\documentclass[showpacs,twocolumn,pre]{revtex4-1}
\usepackage[colorlinks, hyperfigures, linkcolor=black, citecolor=black, breaklinks,]{hyperref}
\usepackage[centertags]{amsmath}
\usepackage{amssymb,eqnarray}
\usepackage{amsthm}
\usepackage{fancyhdr}
\usepackage{graphicx,subfigure}
\usepackage{epsfig}
\usepackage{epsfig}
\usepackage{amssymb}
\usepackage{bbm}
\usepackage{graphicx}
\usepackage{amsmath}
\usepackage{subfigure}

\usepackage{bbold}
\usepackage[arrow]{hhtensor}

\usepackage{epstopdf}

\newcommand{\F}{\boldsymbol{F}}

\newcommand{\kk}{\boldsymbol{k}}

\newcommand{\J}{\vec{J}}
\newcommand{\D}{\Bar{\Bar{\boldsymbol{D}}}}

\newcommand{\rr}{\boldsymbol{r}}

\begin{document}

\title{Effect of anisotropic diffusion on spinodal decomposition}
\author{Hidde Derk Vuijk,\textit{$^{a}$} Joseph Michael Brader,\textit{$^{b}$} and Abhinav Sharma$^{\ast}$\textit{$^{a}$}}
\address{\textsuperscript{a}Leibniz-Institut f\"{u}r Polymerforschung Dresden,
		Institut Theorie der Polymere, 01069 Dresden, Germany\\
		\textsuperscript{b} Universit\'e de Fribourg, Chemin du Mus\'ee 3, CH 1700, Fribourg, Switzerland\\
		}

\begin{abstract}
We study the phase transition dynamics of 
a fluid system in which the particles diffuse anisotropically in space. 
The motivation to study such a situation is provided 
by systems of interacting magnetic colloidal particles subject to the Lorentz force. 
The Smoluchowski equation for the many-particle probability distribution then aquires 
an anisotropic diffusion tensor. We show that anisotropic diffusion results in qualitatively different dynamics of spinodal decomposition compared to the isotropic case. Using the method of dynamical density functional theory, we predict that the intermediate-stage decomposition dynamics are slowed down significantly by anisotropy; the coupling between 
different Fourier modes is strongly reduced. Numerical calculations are performed for a model (Yukawa) fluid that exhibits gas-liquid phase 
separation. 
 \\

\end{abstract}

\maketitle


\footnotetext{\textit{$^{a}$~Leibniz Institute for Polymer Research, Hohe Str. 6, 01069 Dresden, Germany; E-mail: sharma@ipfdd.de}}
\footnotetext{\textit{$^{b}$~University of Fribourg, Chemin du Mus\'ee 3, CH-1700 Fribourg, Switzerland. }}




On sufficiently long time scales, particles suspended in a solvent exhibit random, 
Brownian motion. For interacting systems subject to external force fields the Smoluchowski 
equation provides a complete statistical description of the particle motion and serves as 
a natural generalization of the Einstein diffusion equation. 
In the absence of solvent hydrodynamics and when both the interaction and external forces 
are conservative, the diffusion coefficient entering the 
Smoluchowski equation is a scalar quantity. 
A case of special interest, which has received only limited attention,\cite{balakrishnan2008elements}
is that of magnetic colloids 
subject to the Lorentz force arising from an external magnetic field. 
The application of a magnetic field generates unusual nonequilibrium steady states  
of a quite different character from those generated by other nonconservative driving forces 
(e.g.~shear), which input energy to the system. 
Although the Lorentz force generates particle 
currents, these are purely rotational and no work is done on the system.
For such magnetic systems the Smoluchowski equation picks up a tensorial diffusion coefficient, 
which reflects the anisotropy of the particle motion. The Smoluchowski equation for the positional probability density $P(\rr,t)$ reads as \cite{balakrishnan2008elements}
\begin{align}
\frac{\partial P}{\partial t} = D_{ij}\frac{\partial^2 P}{\partial x_i \partial x_j},
\end{align}
where $x_i$ stands for a Cartesian component of the position of the particle and $D_{ij}$ denotes the $ij^{\text{th}}$ 
component of the diffusion tensor. For a magnetic field $\vec{B} = B\vec{n}$, the diffusion tensor is given by
\begin{align}
D_{ij} = \frac{k_BT}{m\gamma}\left[n_i n_j + \frac{\gamma^2}{\gamma^2 + \omega^2}\left(\delta_{ij} - n_i n_j\right)\right],
\end{align}
where $k_B$ is the Boltzmann constant, $T$ is the temperature, $m$ is the mass of the particle, $\gamma$ is the 
friction coefficient, $q$ is the charge, $\omega = qB/m$ is the cyclotron frequency, and $k_BT/m\gamma$ is the 
isotropic diffusion rate in the absence of magnetic field.
Only the diffusion rate perpendicular to the direction of the  magnetic field
decreases as the field strength increases. The diffusion along the direction of 
magnetic field is unaffected.
Although the structure of the diffusion tensor is well known, \cite{balakrishnan2008elements}
the influence of this on the collective behaviour of interacting systems, 
most notably the phase
transition dynamics, remains to be fully investigated.   

Classical density functional theory (DFT) is a powerful tool for studying the equilibrium structure 
and thermodynamics of interacting, inhomogeneous fluids in external fields.\cite{evans1992density}
The approximate extension of DFT to systems out of equilibrium, dynamical density functional theory 
(DDFT), provides a framework to study the dynamics of overdamped Brownian 
particles.\cite{evans1979nature,dieterich1990nonlinear,marconi1999dynamic,marconi2000dynamic}
The central quantity of interest in DDFT is the ensemble averaged one-body number density, 
$\rho(\rr,t)$, of particles at spatial location $\rr$ and at time $t$. 
DDFT has been used to address the phase transition dynamics which occur during the approach to equilibrium 
from a nonequilibrium initial state. The dynamics of (isotropic) spinodal decomposition at early and 
intermediate times was studied in Ref.~\cite{archer2004dynamical} and the time evolution of 
solidification fronts during crystallization was addressed in Ref.~\cite{archer2014solidification}
Modifications of DDFT to incorporate external driving forces have proven successful in describing the 
phenomenology of some nonequilibrium steady-states; for example, the laning transition in driven colloidal systems.\cite{chakrabarti2003dynamical,brader2011density}

In this paper, we use DDFT to study the dynamics of spinodal decomposition in systems with anisotropic 
diffusion. 
Following the approach of Ref.~\cite{archer2004dynamical} we obtain an equation to describe the short- and intermediate-time dynamics of density fluctuations following a quench into the two-phase region. 
At short times, Fourier components of density fluctuations within a certain range of wavenumbers 
grow exponentially, with the rate of growth in different spatial directions occuring on different 
time scales. 
The intermediate-time dynamics are impacted in a more subtle way that is qualitatively different from the case of isotropic diffusion. Here the coupling of different 
Fourier components allows even those fluctuations to grow that correspond to wavenumbers outside 
the region of linear instability. 
Anisotropic diffusion strongly reduces the coupling between different Fourier components, even for 
small anisotropy, leading to much slower growth of Fourier components with wavenumbers outside 
the region of linear instability. 

The paper is organized as follows. In Sec. \ref{ddft}, we briefly describe the dynamical density theory. We use DDFT with an anisotropic diffusion tensor in Sec. \ref{spinodal_decomposition} to derive an equation for the time evolution of density fluctuations. In Sec. \ref{results} we numerically solve the equation derived in Sec. \ref{spinodal_decomposition} and study the impact of small and large anisotropy on spinodal decomposition. 
Finally we present conclusions and an outlook in Sec. \ref{conclusions}.

\section{Dynamical density functional theory}\label{ddft}
The time evolution of the density distribution $\rho(\rr,t)$ in a fluid system of particles is given by the continuity equation
\begin{align}
 \frac{\partial \rho(\rr,t)}{\partial t} = - \nabla \cdotp \J,
 \label{continuity}
\end{align}
where $\J$ is the current of particles. This equation merely expresses the fact that the number of particles is conserved in the system. 
Although formal expression for the current can be derived, it is often necessary to make approximations to find solutions to this equation. Dynamical density functional theory  provides such an approximation for $\J$ which has been found to be rather accurate.\cite{marconi1999dynamic,marconi2000dynamic}  
DDFT approximates the current $\J$ as
\begin{align}
\J = -\beta \D \cdot \rho(\rr,t) \nabla \frac{\delta \F[\rho]}{\delta \rho(\rr,t)},
\label{flux}
\end{align}
where $\beta = 1/(k_B T)$, $\D$ is an (arbitrary) diffusion tensor and $\F$ is the equilibrium Helmholtz free energy functional. 
In the absence of external potential, the Helmholtz free energy functional is given as follows:
\begin{align}
\F[\rho(\rr,t)] = k_B T \int d\rr \rho(\rr,t)\left[\ln (\rho(\rr,t) \Lambda^3)- 1\right] + \F_{\mathrm{ex}}[\rho(\rr,t)].
\label{functional}
\end{align}
The first term is the ideal gas free energy and $\Lambda$ is the thermal wavelength. 
$\F_{\mathrm{ex}}[\rho]$ is the excess free energy; that is, the contribution due to the interactions between the particles.

The main assumption underlying DDFT is that the correlations between the particles when the fluid is out of equilibrium are
the same as in an equilibrium fluid with the same one-body density profile $\rho(\rr,t)$. 
This is a major assumption which cannot be justified \emph{a priori}. 
Nevertheless, DDFT has proven highly successful in describing the approach of a system towards equilibrium from a nonequilibrium initial state.\cite{archer2014solidification}
It is remarkable that DDFT can successfully describe the steady states of systems that are continuously driven out of equilibrium, such as sheared colloidal 
suspensions \cite{chakrabarti2003dynamical,brader2011density} and active Brownian particles. \cite{menzel2016dynamical,sharma2017brownian}

The functional derivative in Eq. \eqref{flux} can be interpreted as the chemical potential $\mu(\rr,t)$ acting on a particle located at ($\rr,t)$. 
One can then calculate the net driving force on the particle as the spatial gradient of the chemical potential, $\nabla \mu(\rr,t)$, and obtain the current $\J = -\beta \D \cdot \rho(\rr,t) \nabla \mu(\rr,t)$. The chemical potential obtained from Eq. \eqref{flux} has two contributions
\begin{align}
\frac{\delta \F[\rho]}{\delta \rho(\rr)} \equiv \mu(\rr,t)= \mu_{\mathrm{id}} + \mu_{\mathrm{ex}},
\label{fderiv}
\end{align}
where
\begin{align}
\mu_{\mathrm{id}} = k_B T\ln \left[\rho(\rr) \Lambda^3\right]
\end{align}
is the ideal gas contribution to the chemical potential, and
\begin{align}
\mu_{\mathrm{ex}} = \frac{\delta \F_{\mathrm{ex}}[\rho]}{\delta \rho(\rr)} \equiv -k_B Tc^{(1)}(\rr),
\label{c1}
\end{align}
is the contriubution to the chemical potential comming from the interactions between the particles, where $c^{(1)}(\rr)$ is the one-body direct correlation function.\cite{evans1979nature}
The functional derivative of $c^{(1)}(\rr)$ with respect to $\rho(\rr)$ yields the Ornstein-Zernike pair direct correlation function of the fluid \cite{hansen1990theory}
\begin{align}
k_B T c^{(2)}(\rr, \rr') \equiv -\frac{\delta^2\F_{\mathrm{ex}}[\rho]}{\delta \rho(\rr) \delta \rho(\rr')},
\label{c2}
\end{align}
which is one of the key quantities of interest in liquid state theories.\cite{hansen1990theory}
Below we show how $c^{(2)}(\rr, \rr') $ appears naturally in description of the dynamics of spinodal decomposition.

\section{Spinodal decomposition}\label{spinodal_decomposition}
\begin{figure}[t]
 \centering
 \vspace{-1cm}
 \includegraphics[height=10cm]{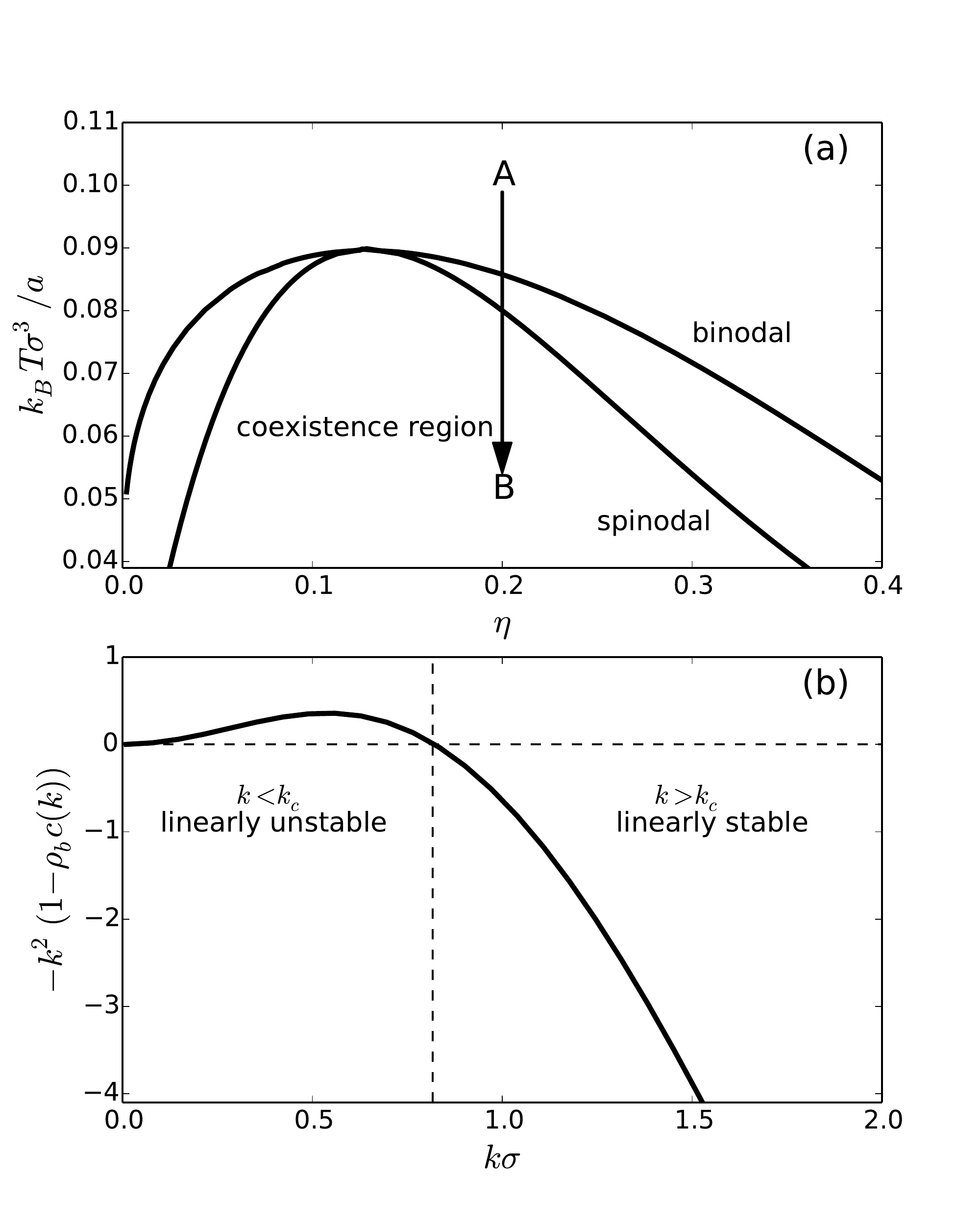}
 \caption{(a) The phase diagram of the fluid system composed of particles interacting via pair potential that is infinitely repulsive for $r <\sigma$ and is attractive (Yukawa) for $r >\sigma$. 
 The packing fraction is denoted as $\eta = \pi\rho_b\sigma^3/6$ where $\sigma$ is the hard-sphere diameter of a particle and $\rho_b$ is the bulk number density. 
 The parameter $k_B T \sigma^3/a$ is the reduced temperature. The parameter $a$ governs the energy scale of the attractive interaction between particles (Eq. \eqref{vat}). 
 The arrow shows the quench from state point A to state point B in the coexistence region. 
 The state point B correponds to $k_B T \sigma^3/a = 0.05$ and $\eta = 0.2$. 
 All the results presented below correspond to the dynamics of the phase separation evolving from this initial state point.
 (b) The function $-k^2(1-\rho_b\hat{c}^{(2)}(k))$ is shown for the case of isotropic diffusion.
 This function describes the rate of growth of the density fluctuations in Eq. \eqref{fourier} at short times.
 For $k<k_c\approx 0.8$, where this function is positive, the density fluctuations grow exponentially.
 For $k>k_c$, any fluctuation is damped. The critical wavenumber $k_c$ depends on how far the state point lies inside the coexistence region.
 }
 \label{phasediagram}
\end{figure}

In this section, we apply the DDFT in Sec. \ref{ddft} to fluid spinodal decomposition.
When a colloidal fluid is quenched to a state point inside the binodal, the fluid undergoes phase seperation into, for instance,
a liquid and a gas phase. There are two mechanisms of phase seperation.
The mechanism, referred to as  nucleation occurs when droplets of one phase form in the other phase.\cite{gunton1983dynamics,onuki2002phase} 
Phase separation proceeds by nucleation when the fluid is quenched to state point that lies inside the binodal but close to it. 
This region corresponds to the region lying within the spinodal and binodal lines (see Fig. \ref{phasediagram}). 
The other mechanism of phase seperation is called spinodal decomposition and it occurs when the state point lies well inside the binodal. 
Spinodal decomposition is characterized by the exponential growth of density fluctuations of certain wavelengths.\cite{gunton1983dynamics,onuki2002phase} 
Experimentally, however, there is not a sharp distinction between regions where phase separation occurs via nucleation and via spinodal decomposition.\cite{archer2004dynamical} 
Nevertheless, it is generally accepted that for a deep quench into the coexistence region, the phase seperation occurs via spinodal decomposition. 
In a fluid undergoing spinodal decomposition three different regimes can be distinguished. For early times after the quench, the amplitude of the density fluctuations are small and 
theories linear in the density fluctuations, such as the well-known Cahn--Hilliard theory, \cite{cahn1959free,cahn1961spinodal}  provide a good description of this  early  stage of 
spinodal decomposition. At intermediate times the density fluctuations can be large, but sharp interfaces between domains of 
gaslike and liquidlike regions have still not formed.\cite{dhont1996spinodal} 
At long times sharp interfaces develop between domains of liquid and gas. 
The Allen-Cahn theory explicitly takes into account the dynamics of the interfaces, and successfully describes the dynamics of spinodal decomposition in this regime.\cite{allen1979microscopic}
In this paper, we focus only on the short- and intermediate-time dynamics of spinodal decomposition.

\begin{figure*}[t]
 \centering
 \begin{tabular}{ccccc}
  \hspace{-0.6cm} \includegraphics[width = 0.247\textwidth]{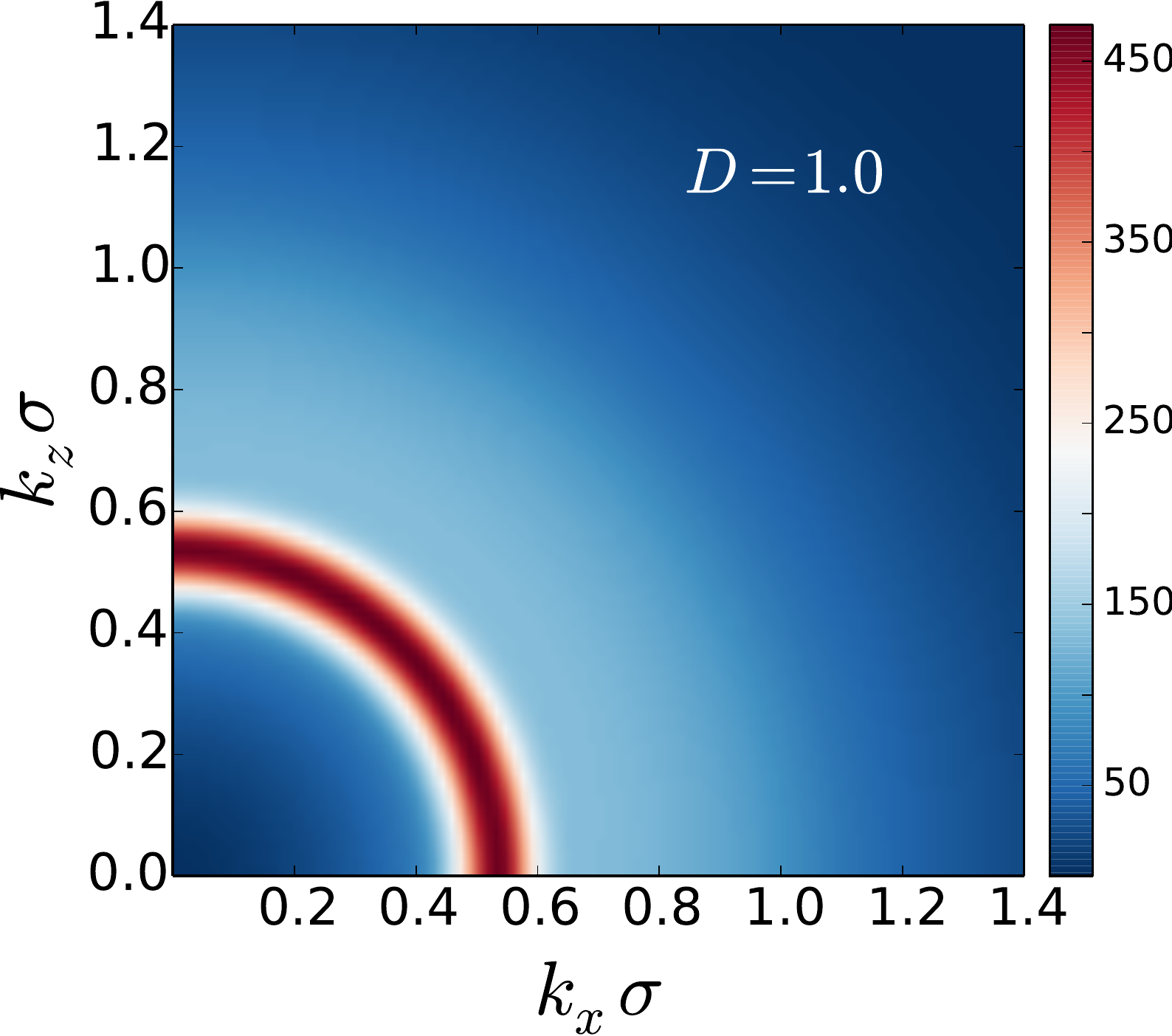} & \hspace{-0.1cm} \includegraphics[width = 0.22\textwidth]{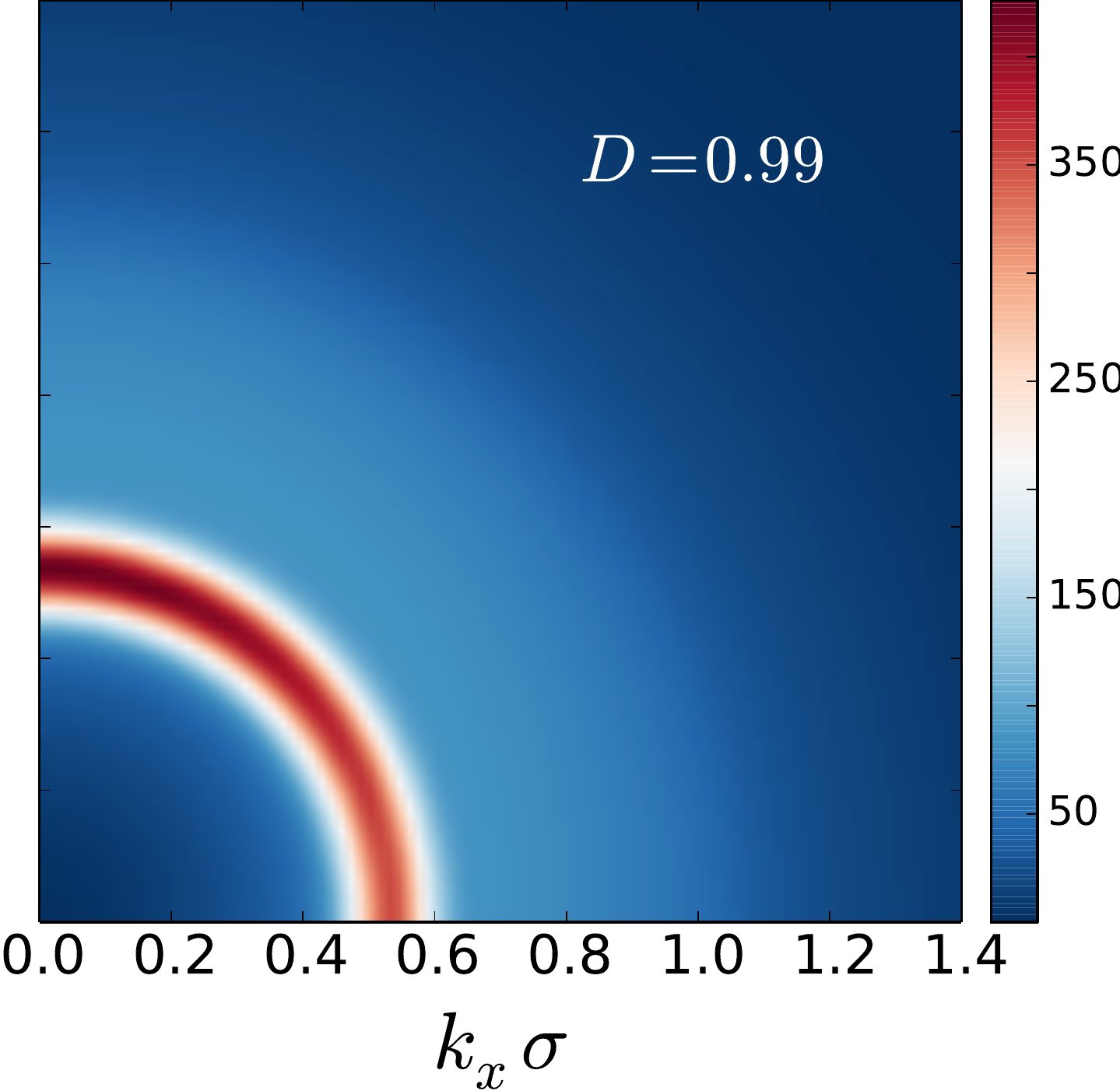} &\hspace{-0.1cm} \includegraphics[width = 0.22\textwidth]{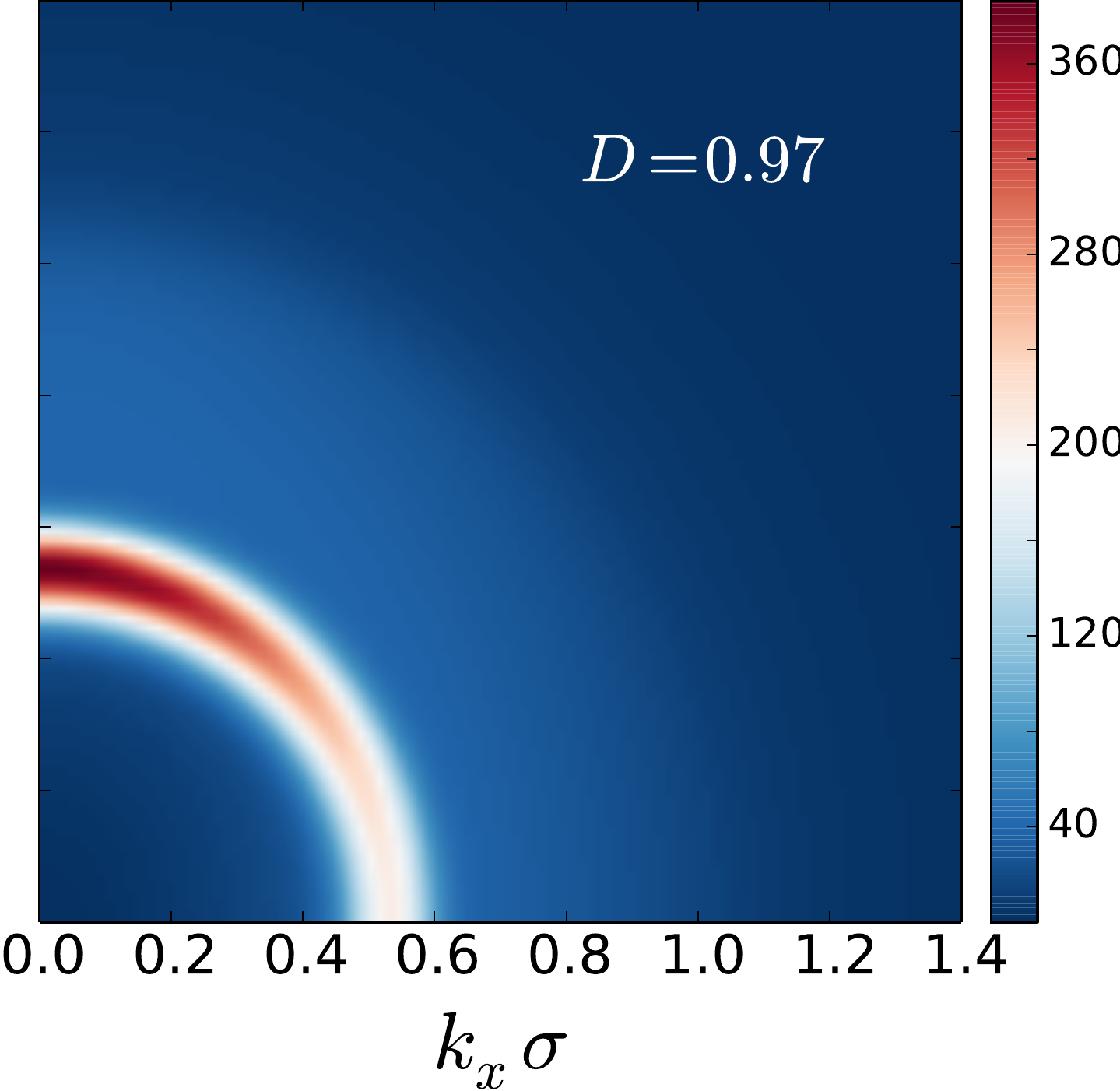} &\hspace{-0.1cm} \includegraphics[width = 0.22\textwidth]{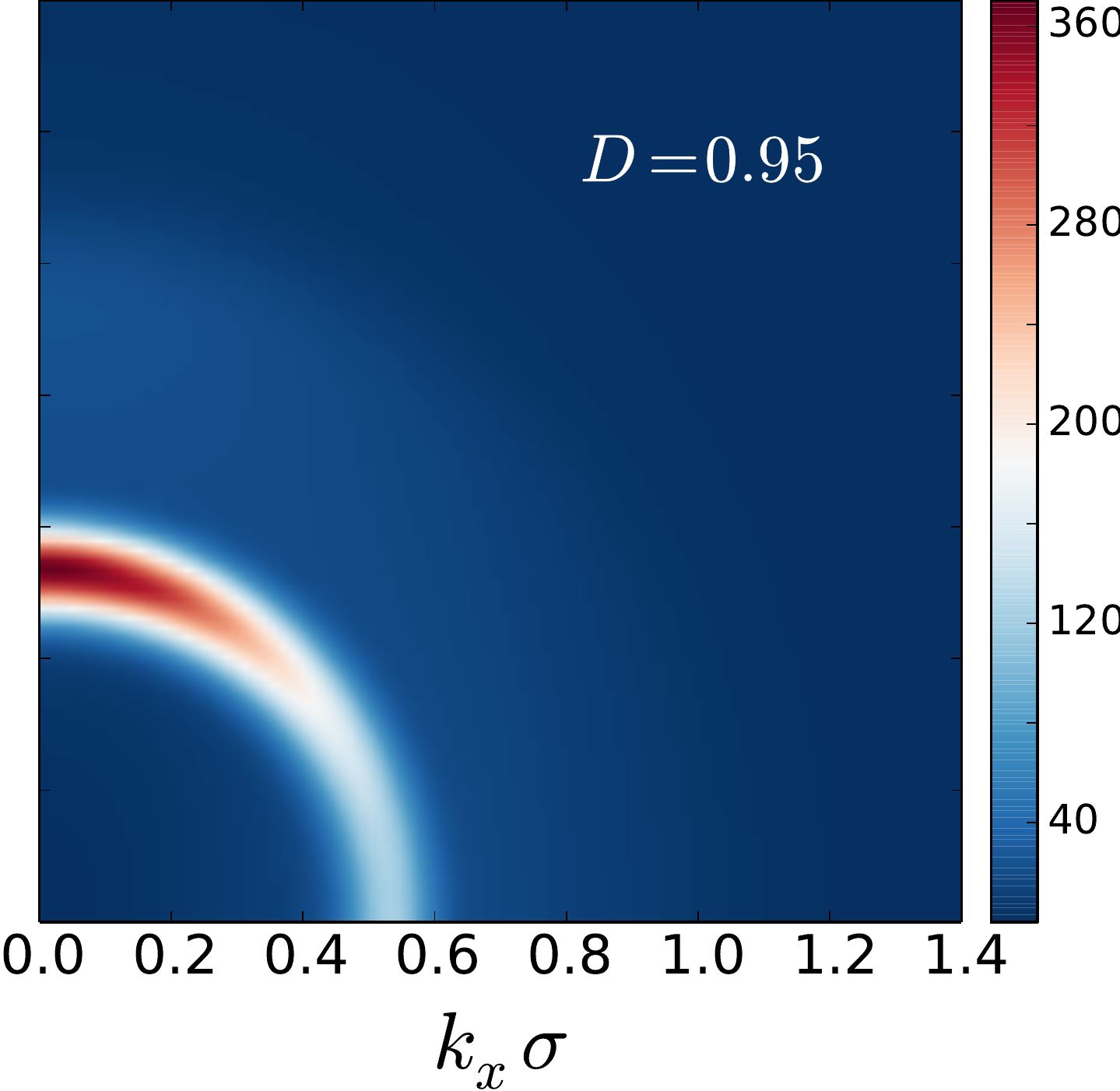} \\ 
    \hspace{-0.6cm} \includegraphics[width = 0.247\textwidth]{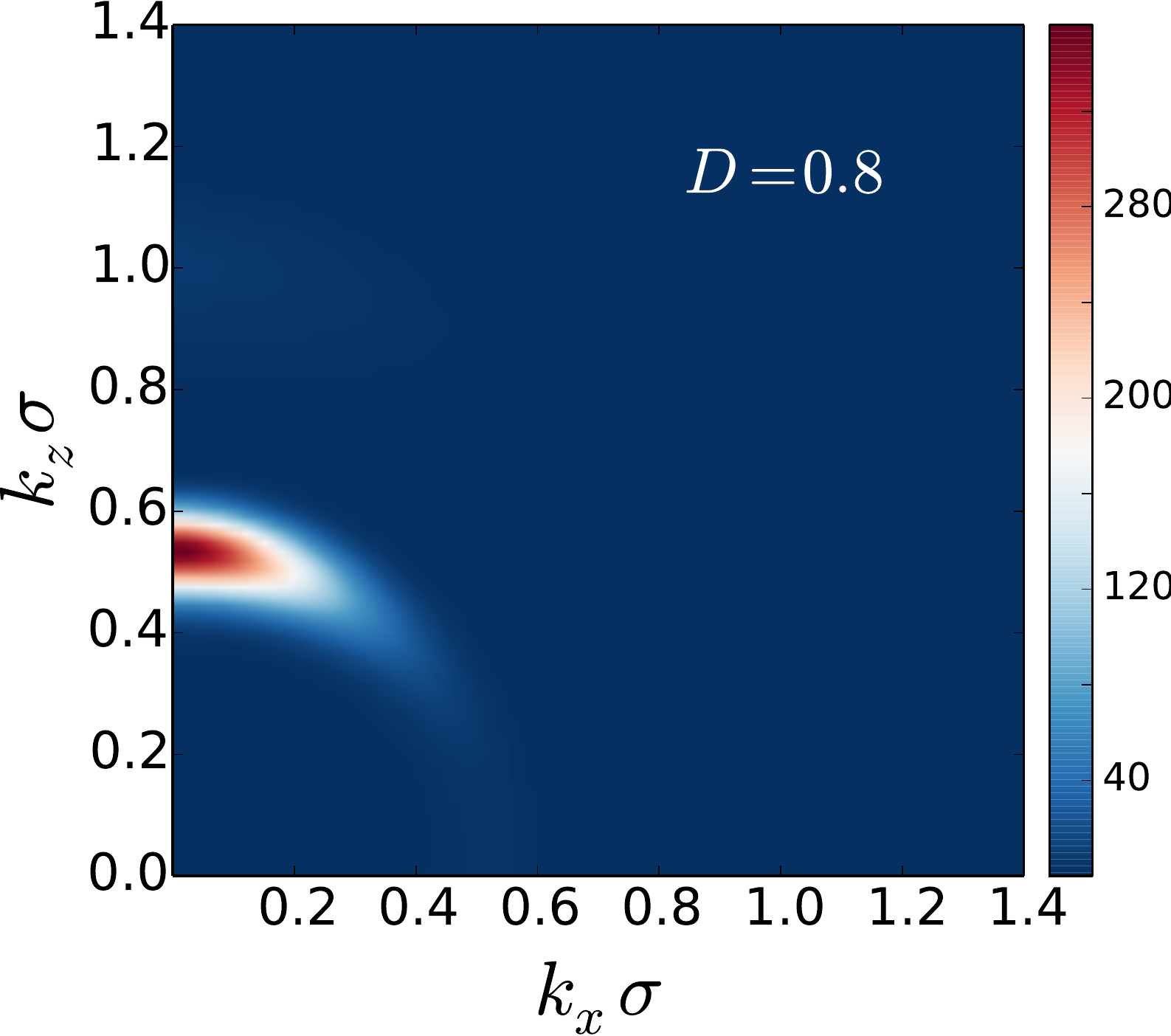} & \hspace{-0.3cm} \includegraphics[width = 0.22\textwidth]{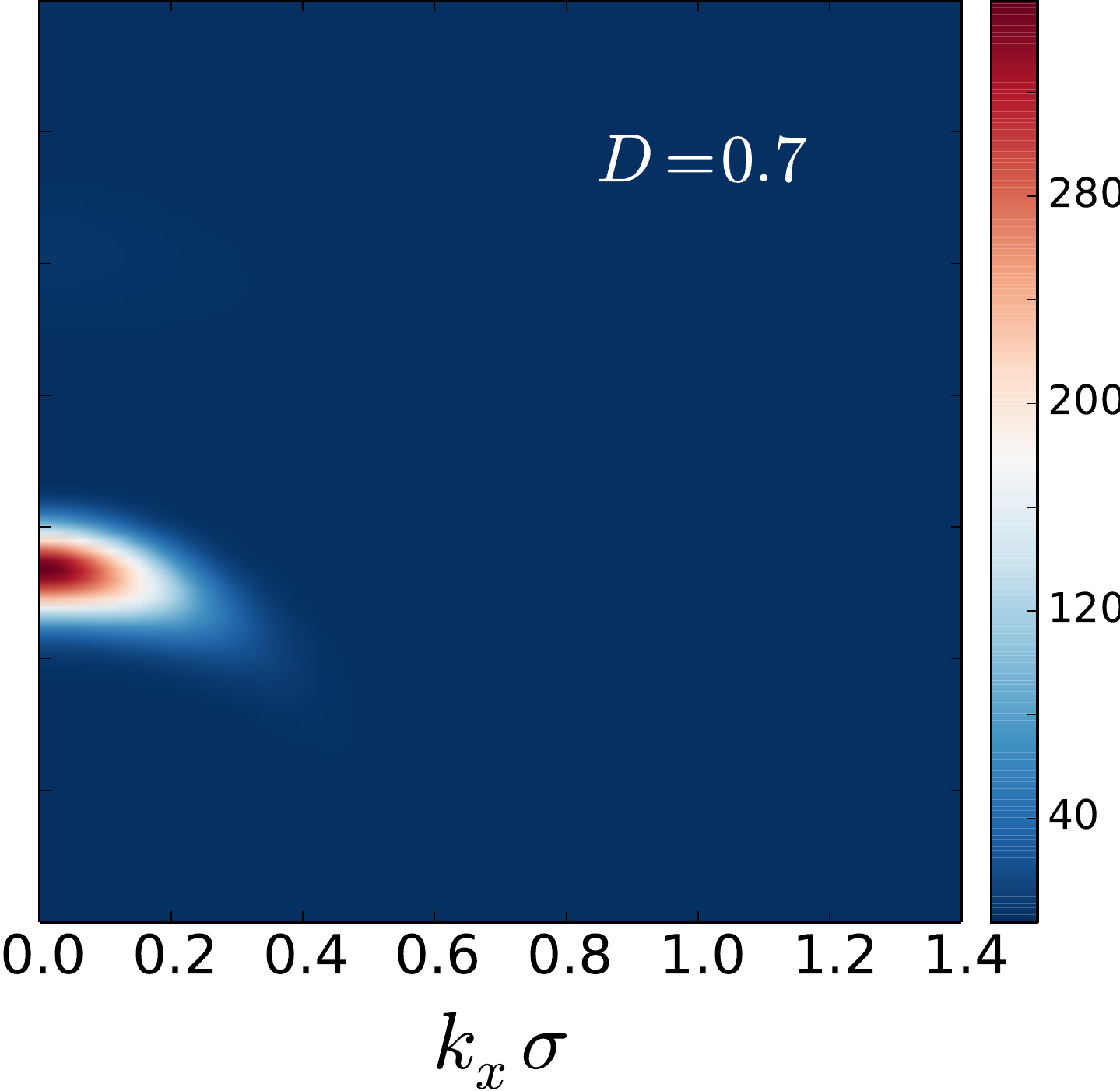} &\hspace{-0.1cm} \includegraphics[width = 0.22\textwidth]{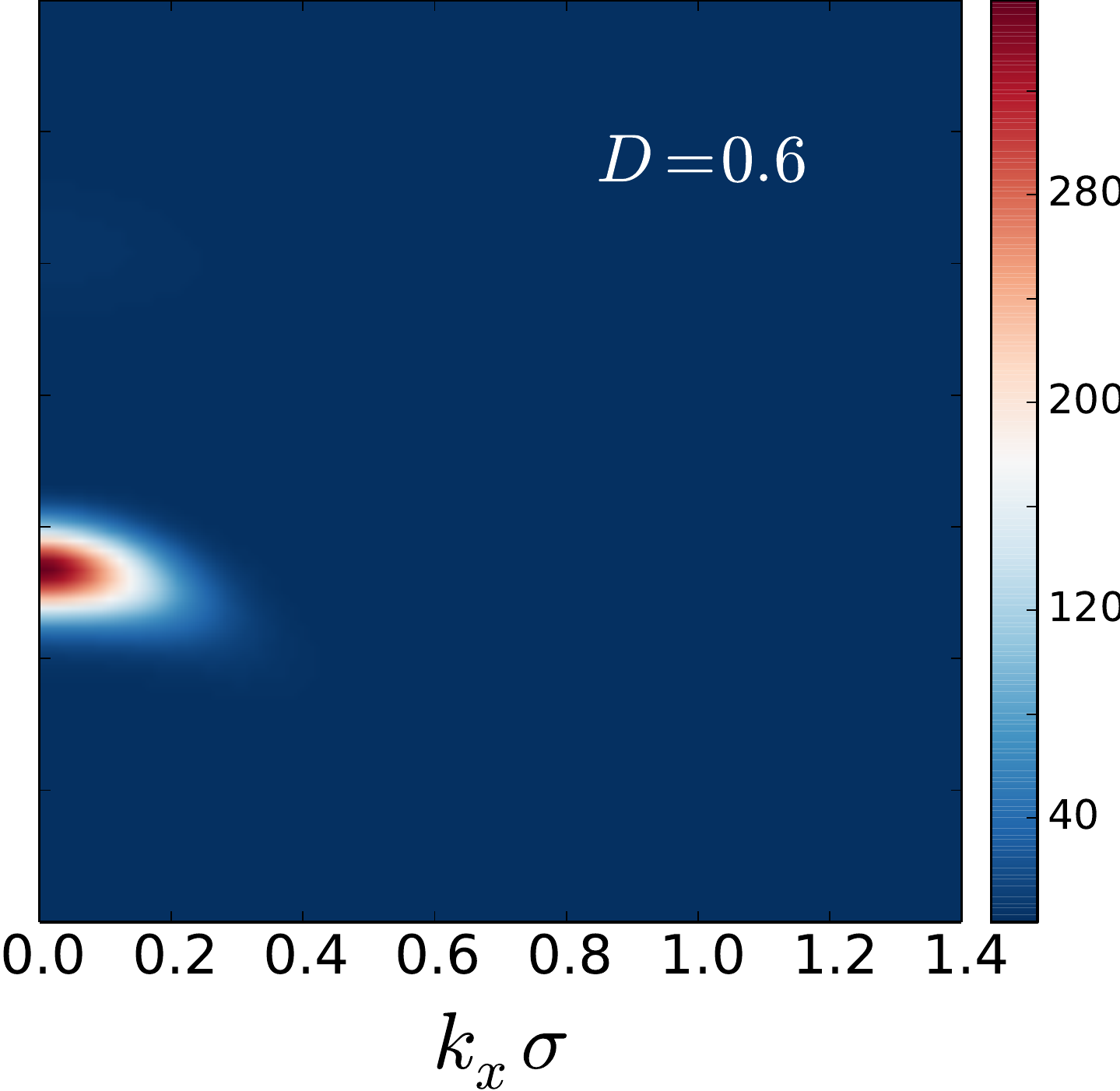} &\hspace{-0.1cm} \includegraphics[width = 0.22\textwidth]{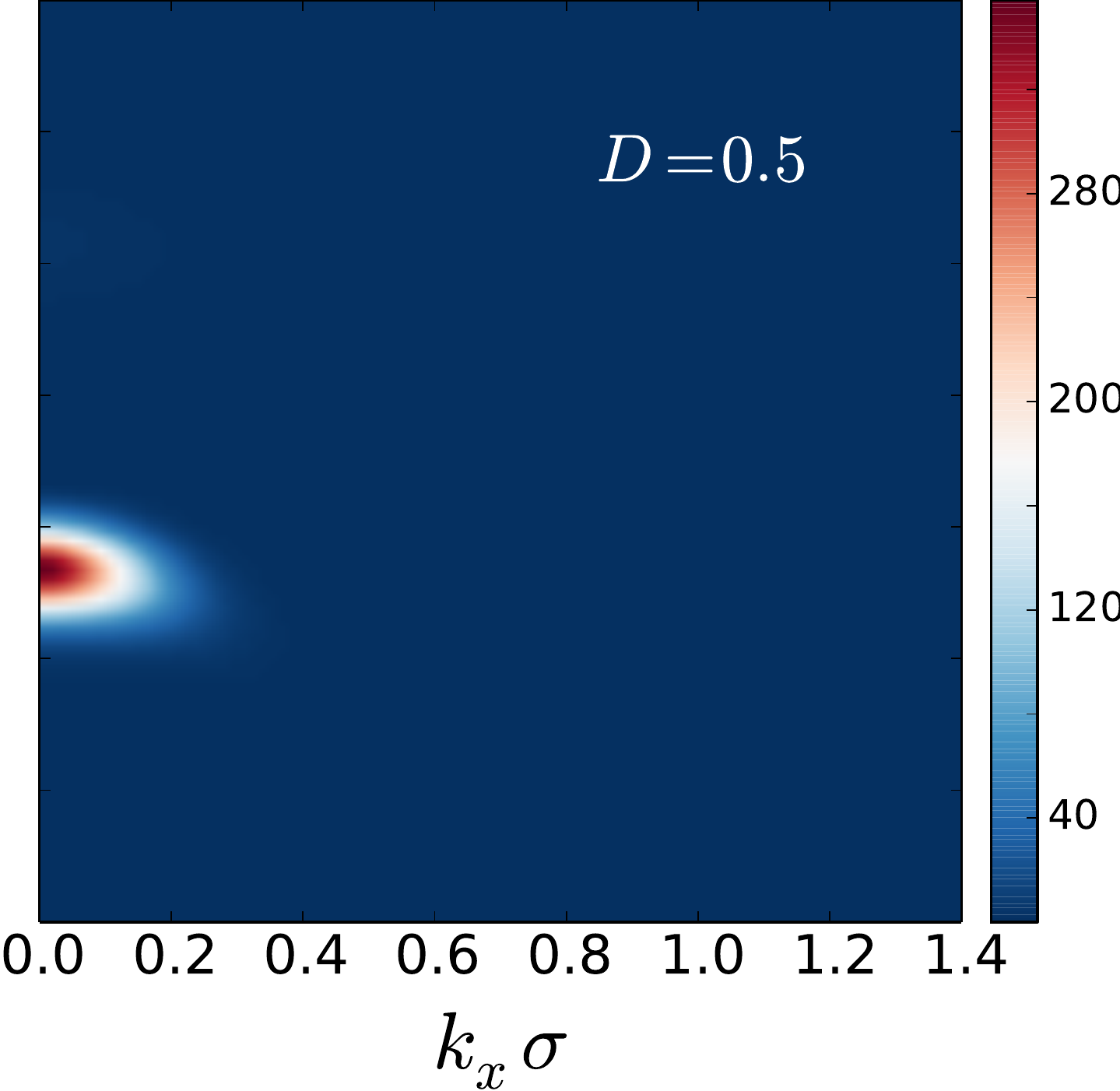}\\
  \hspace{-0.6cm} \includegraphics[width = 0.247\textwidth]{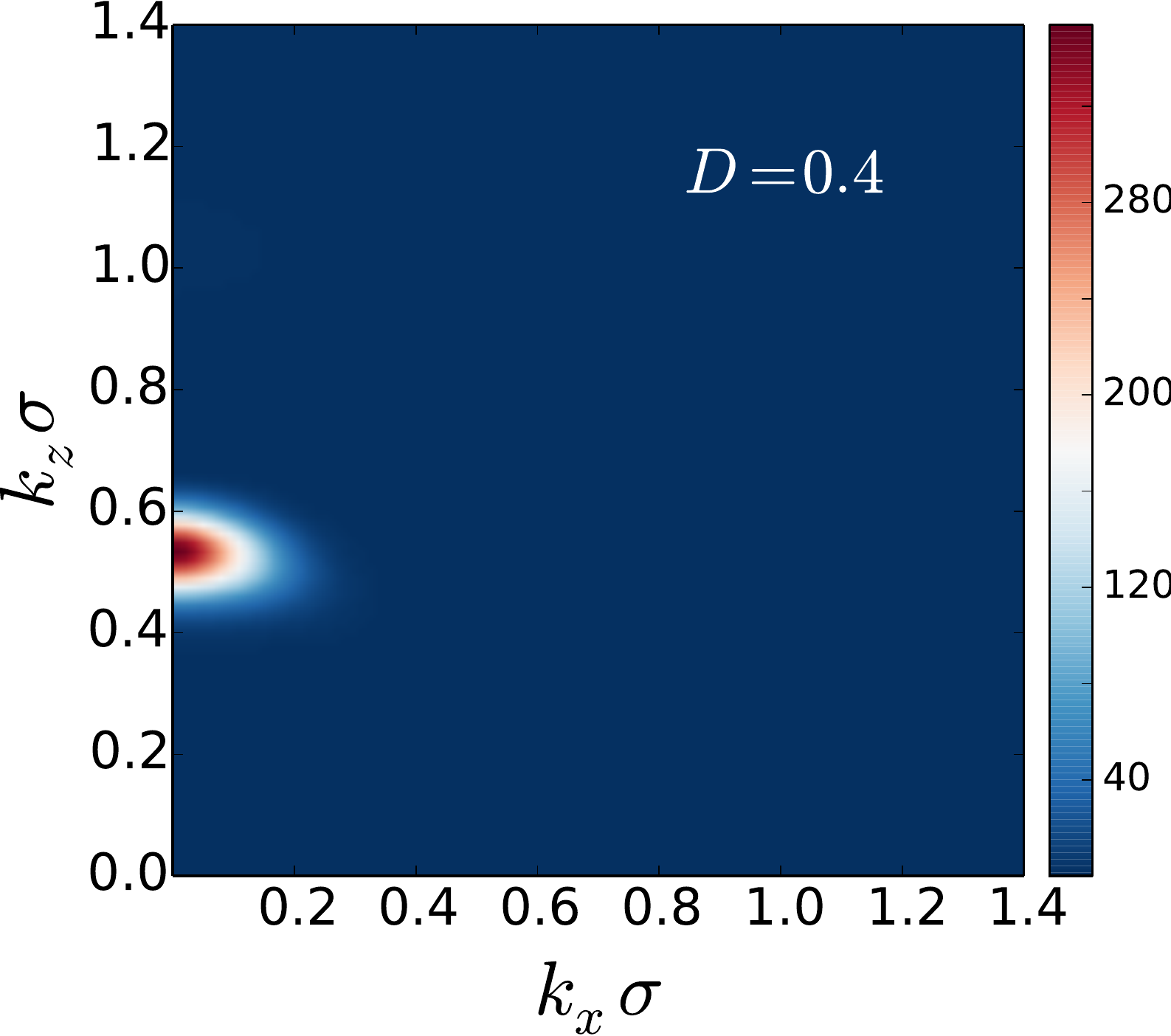} & \hspace{-0.1cm} \includegraphics[width = 0.22\textwidth]{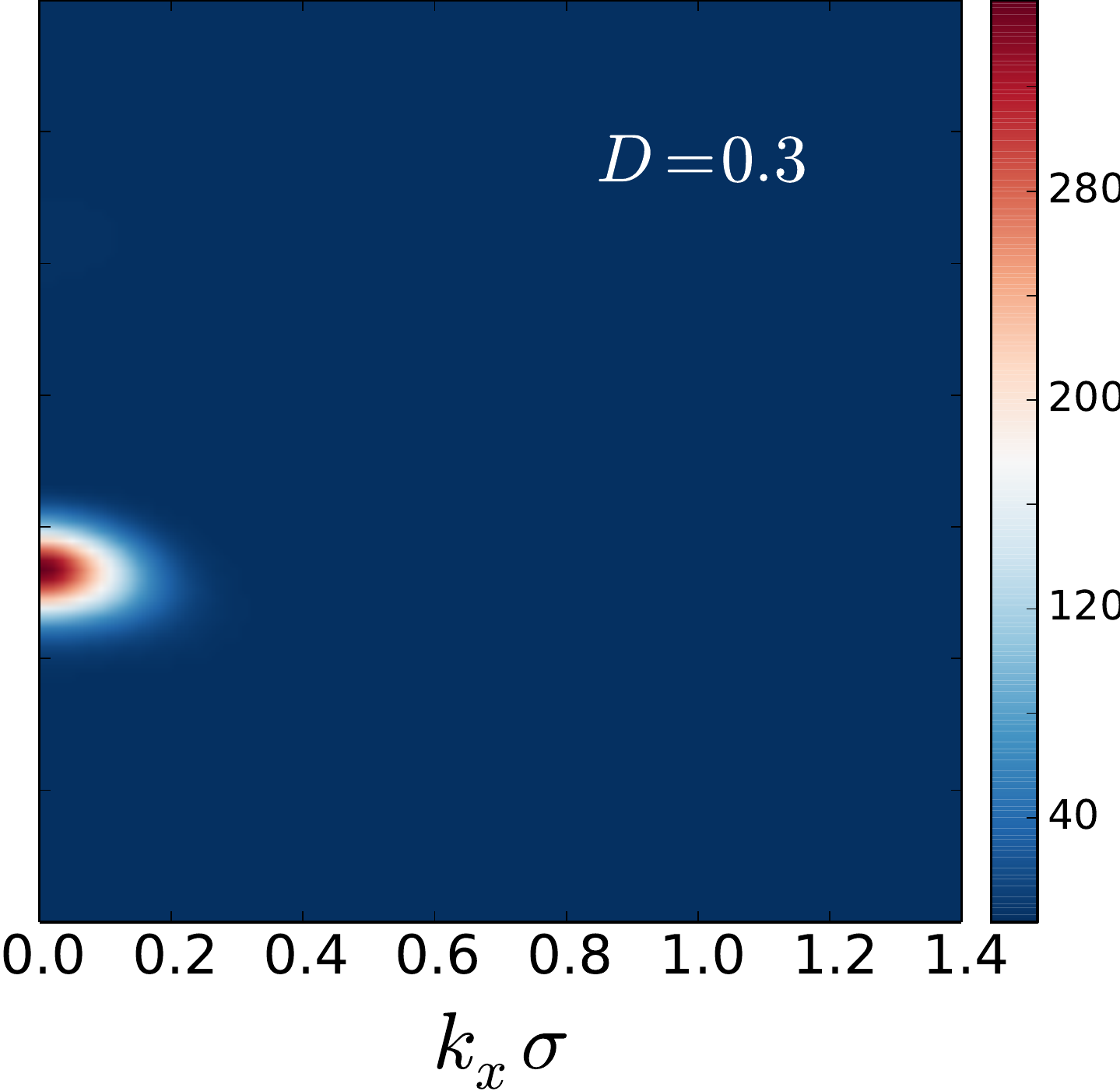} &\hspace{-0.1cm} \includegraphics[width = 0.22\textwidth]{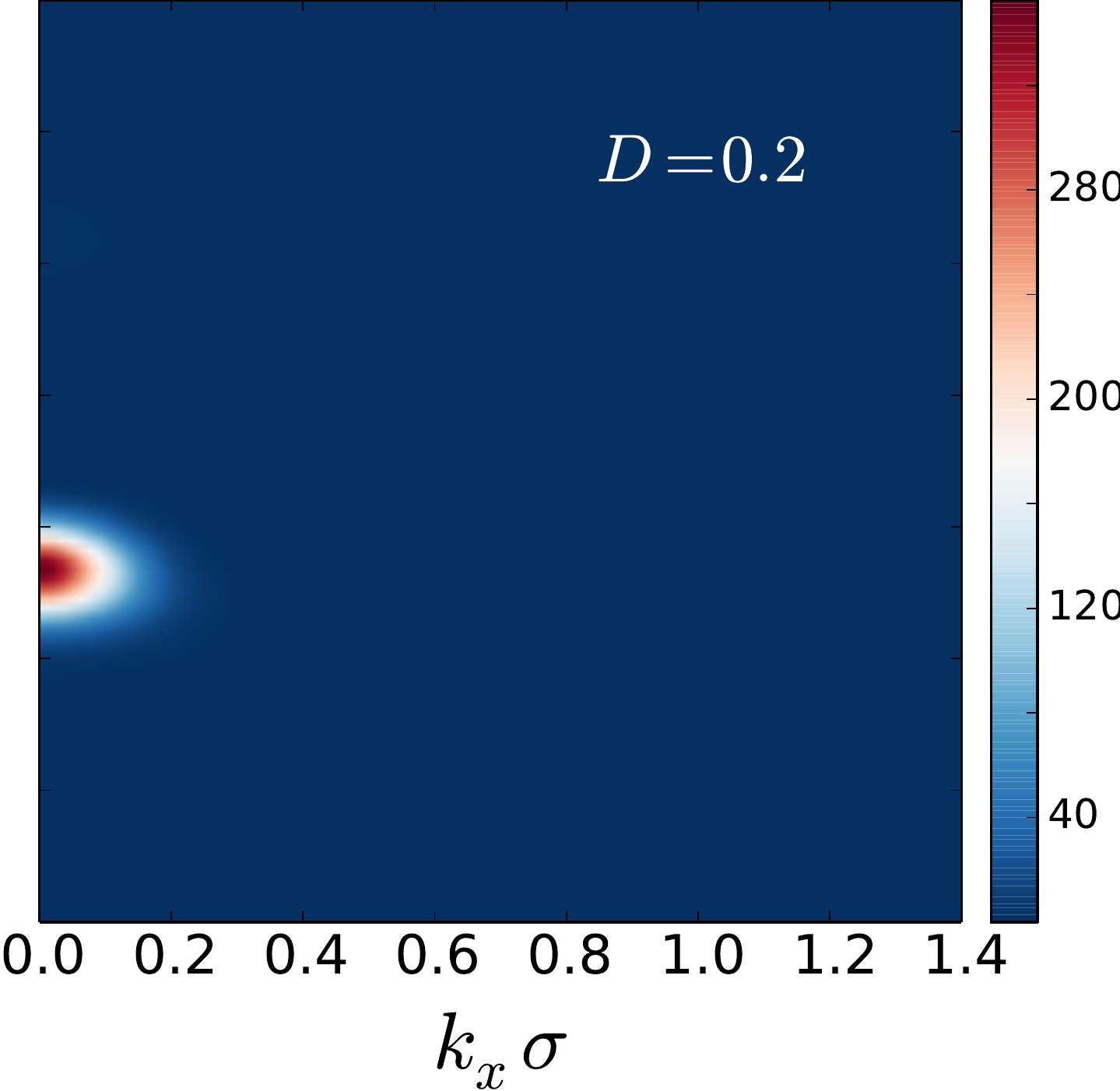} &\hspace{-0.1cm} \includegraphics[width = 0.22\textwidth]{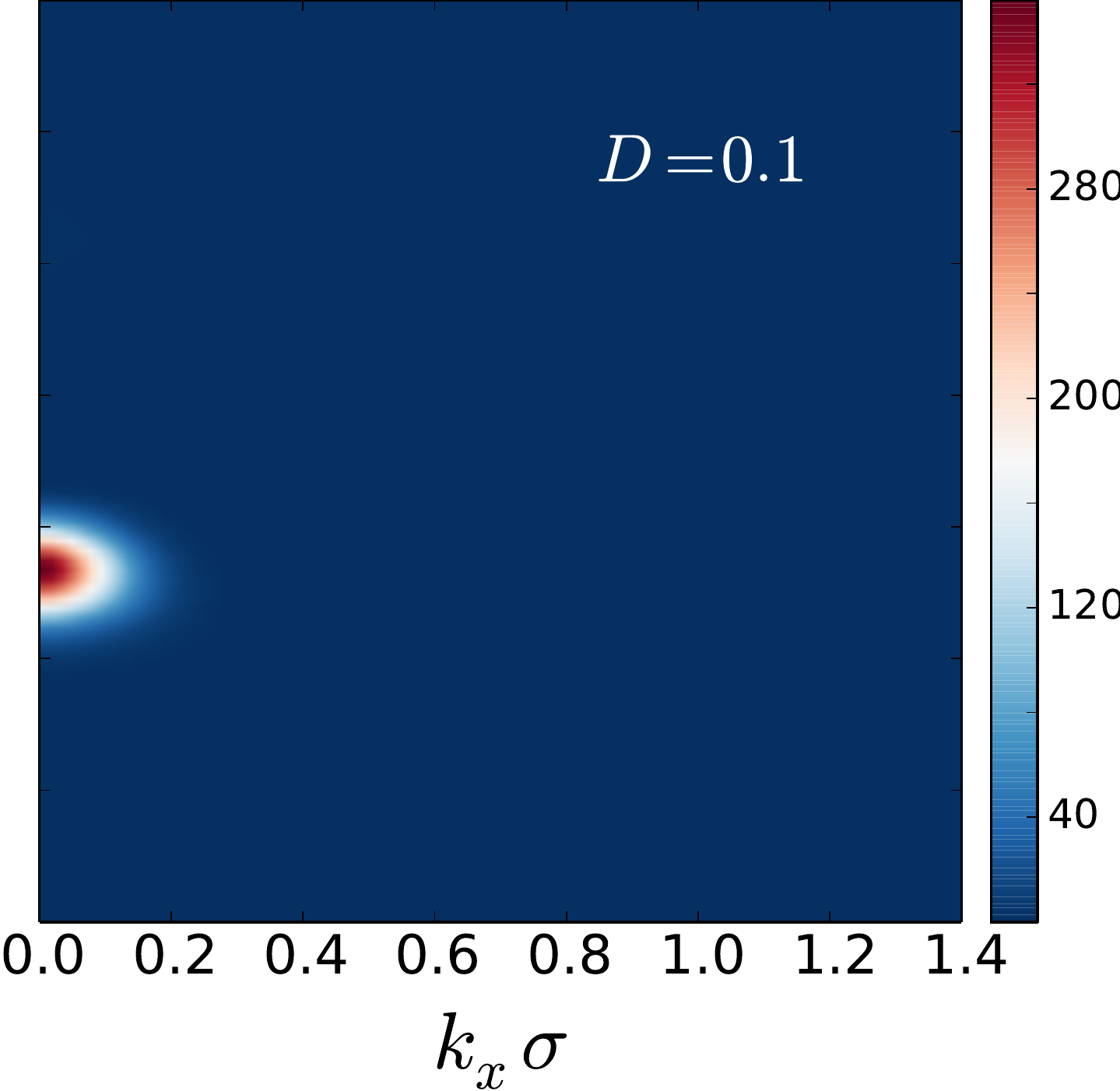}
 \end{tabular}
   \caption{Colorplots of $\hat \rho(\kk)$ shown in the $k_x$-$k_z$ plane at time $\tilde t = \sigma^2 t/D_{zz} = 69$ with $D_{zz}=1$
   for different values of $D\leq1$ (see legend).
   The top row of figures correspond to small anisotropy whereas the bottom row to large anisotropy. 
   In the case of isotropic diffusion $\hat \rho(\kk) = \hat \rho(k)$.
   Fourier components with wavenumbers $k<k_c$ have the largest magnitude. 
   Fourier components outisde the range $k<k_c$ also grow due to the nonlinear coupling between different Fourier components. 
   However, with increasing anisotropy, the growth of the Fourier components outside the $k<k_c$ region becomes increasingly suppressed. 
   When the anisotropy is large ($D<0.5$) only the Fourier components along the $(0,0,k_z)$ direction show any significant growth. 
 }
 \label{kxkzplanes}
\end{figure*}

\begin{figure*}[t]
 \centering
 \begin{tabular}{ccccc}
  \hspace{-0.6cm} \includegraphics[width = 0.247\textwidth]{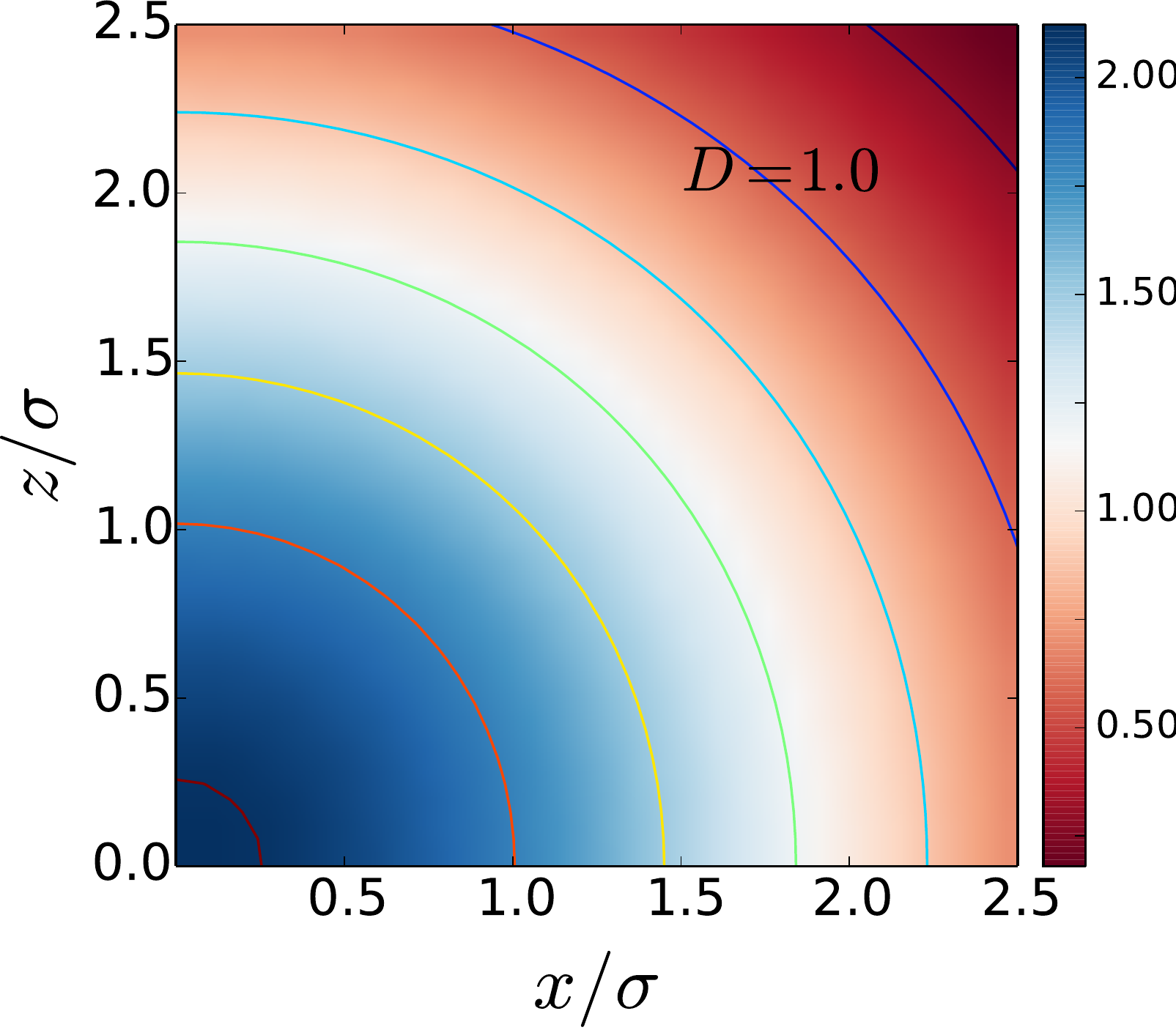} & \hspace{-0.1cm} \includegraphics[width = 0.22\textwidth]{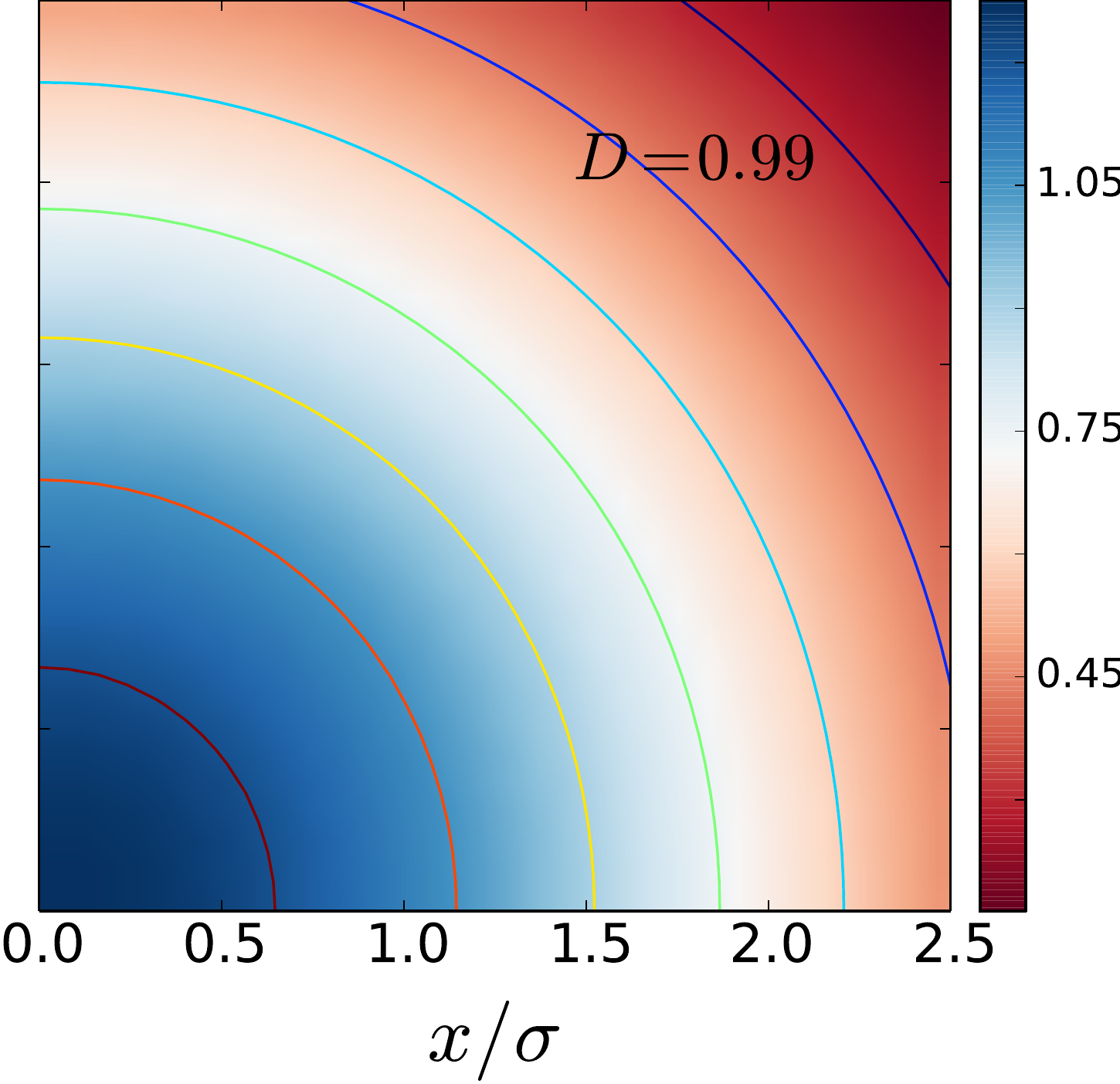} &\hspace{-0.1cm} \includegraphics[width = 0.22\textwidth]{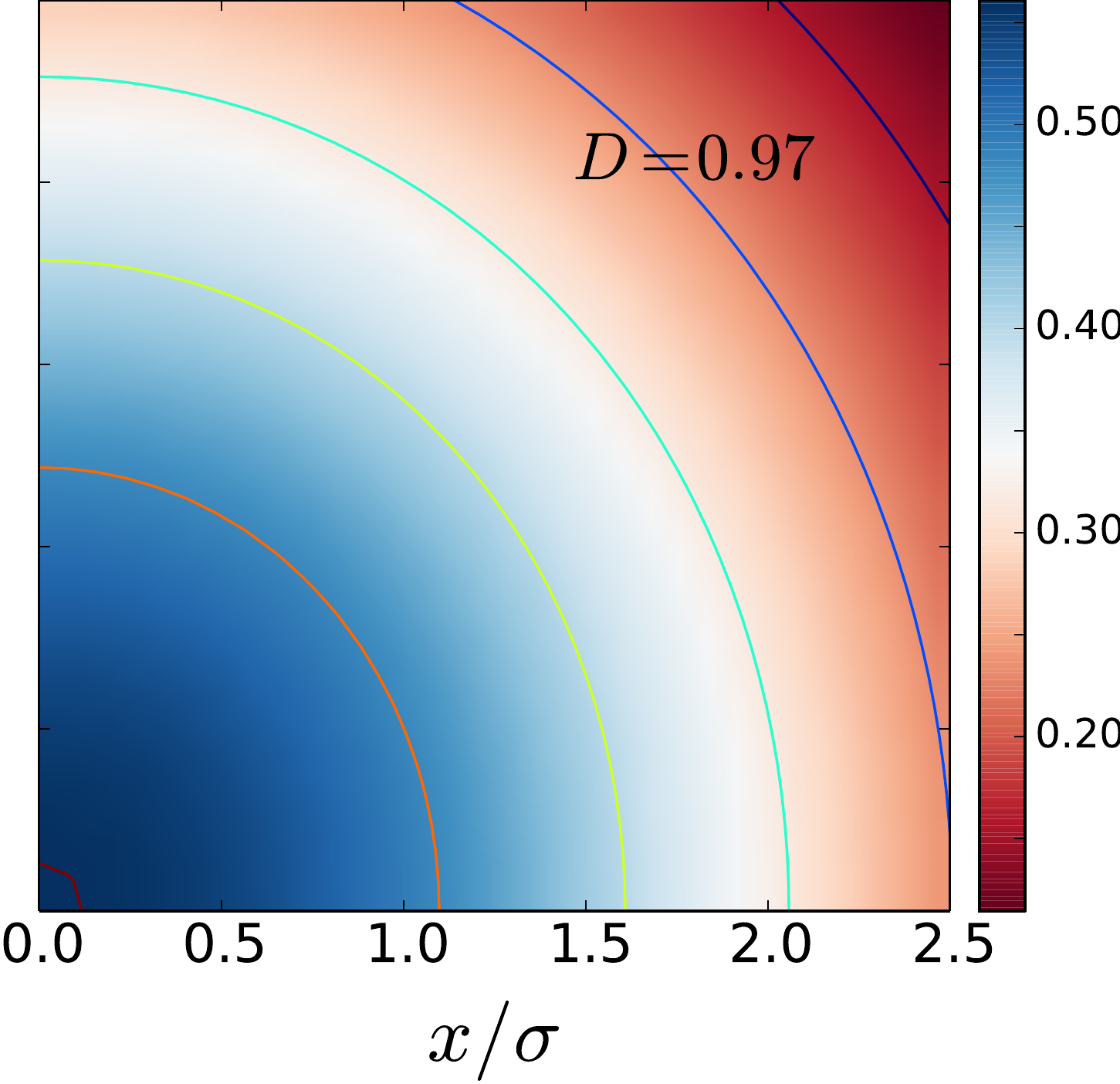} &\hspace{-0.1cm} \includegraphics[width = 0.22\textwidth]{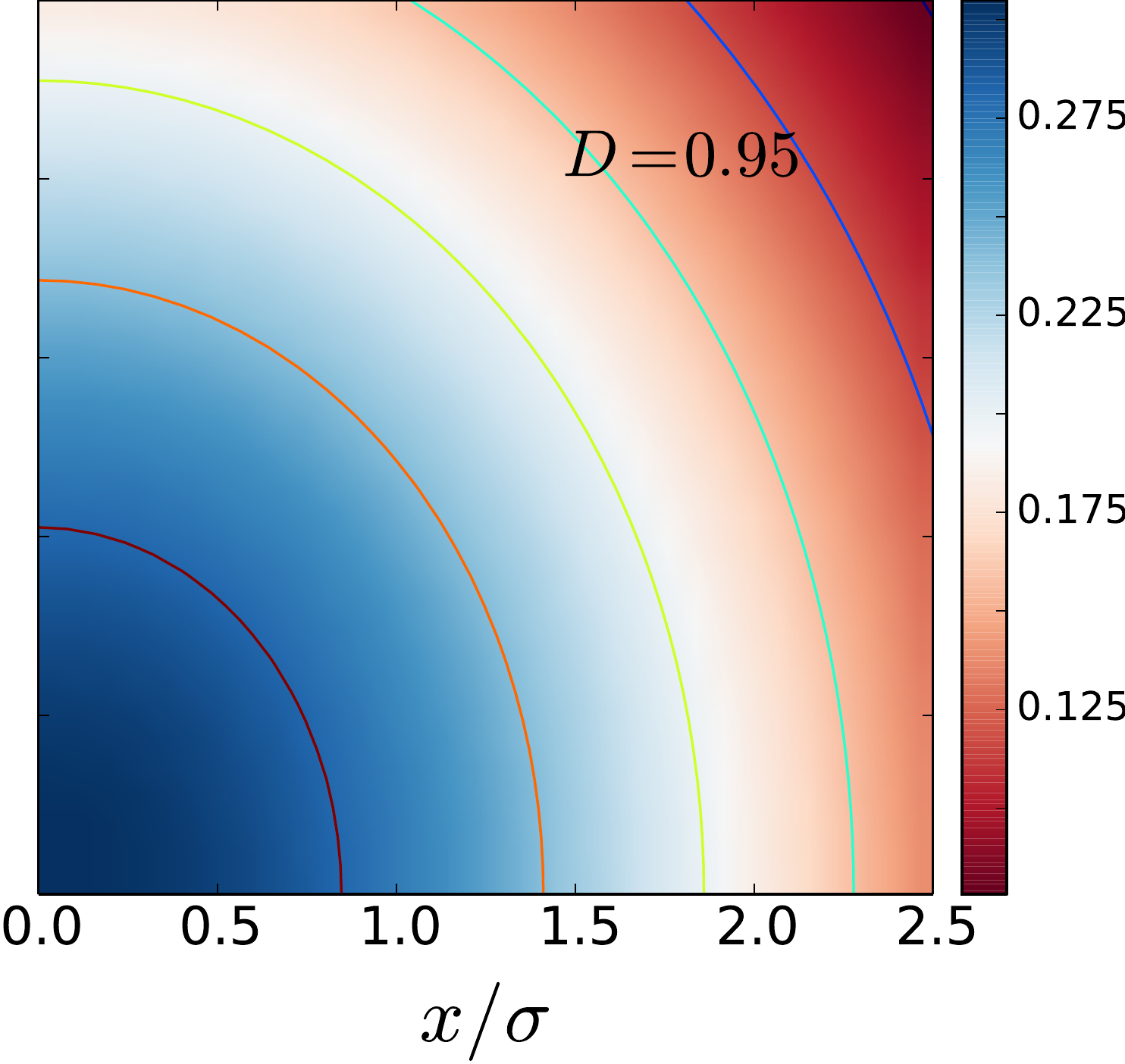} \\ 
    \hspace{-0.6cm} \includegraphics[width = 0.247\textwidth]{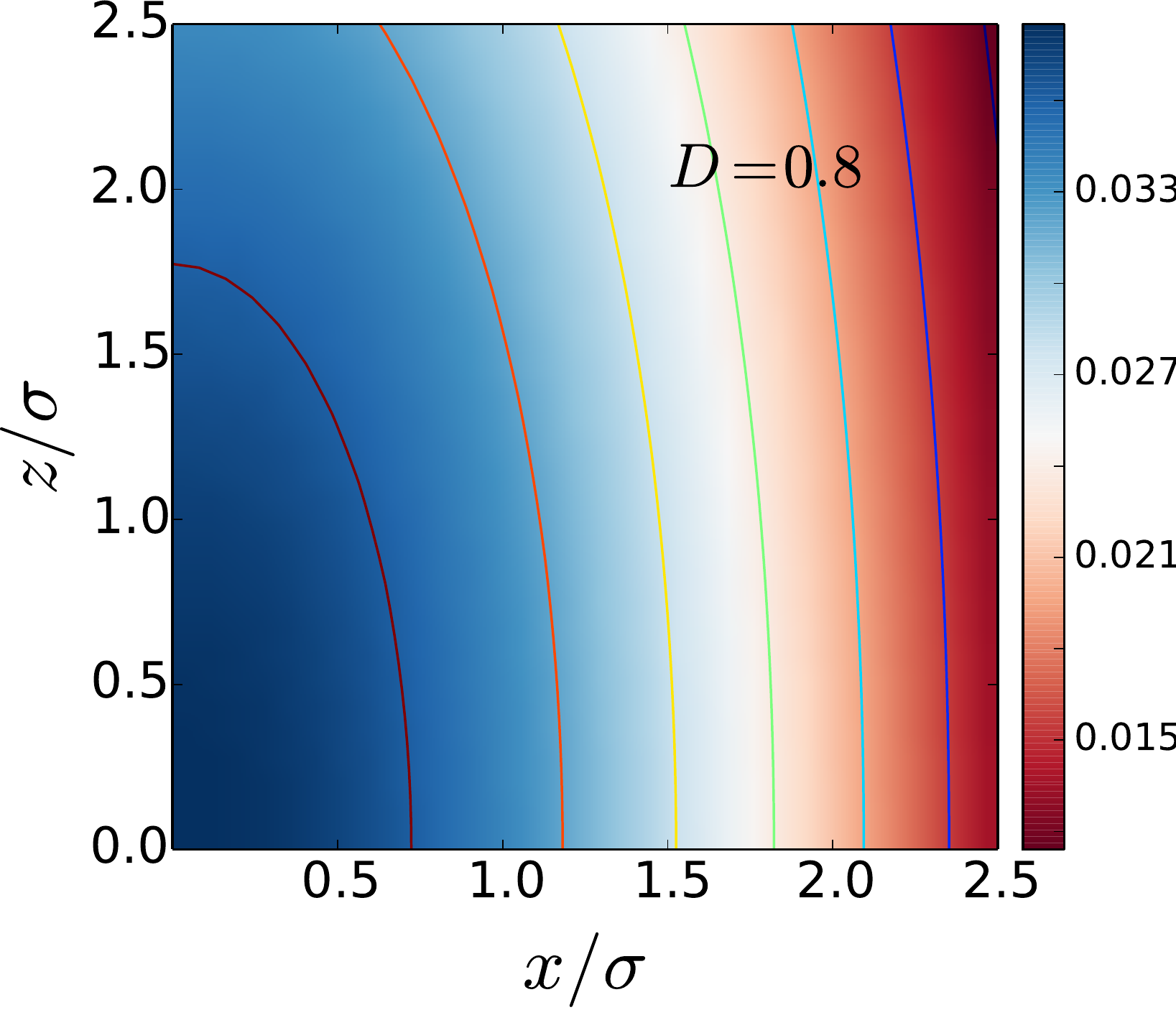} & \hspace{-0.1cm} \includegraphics[width = 0.22\textwidth]{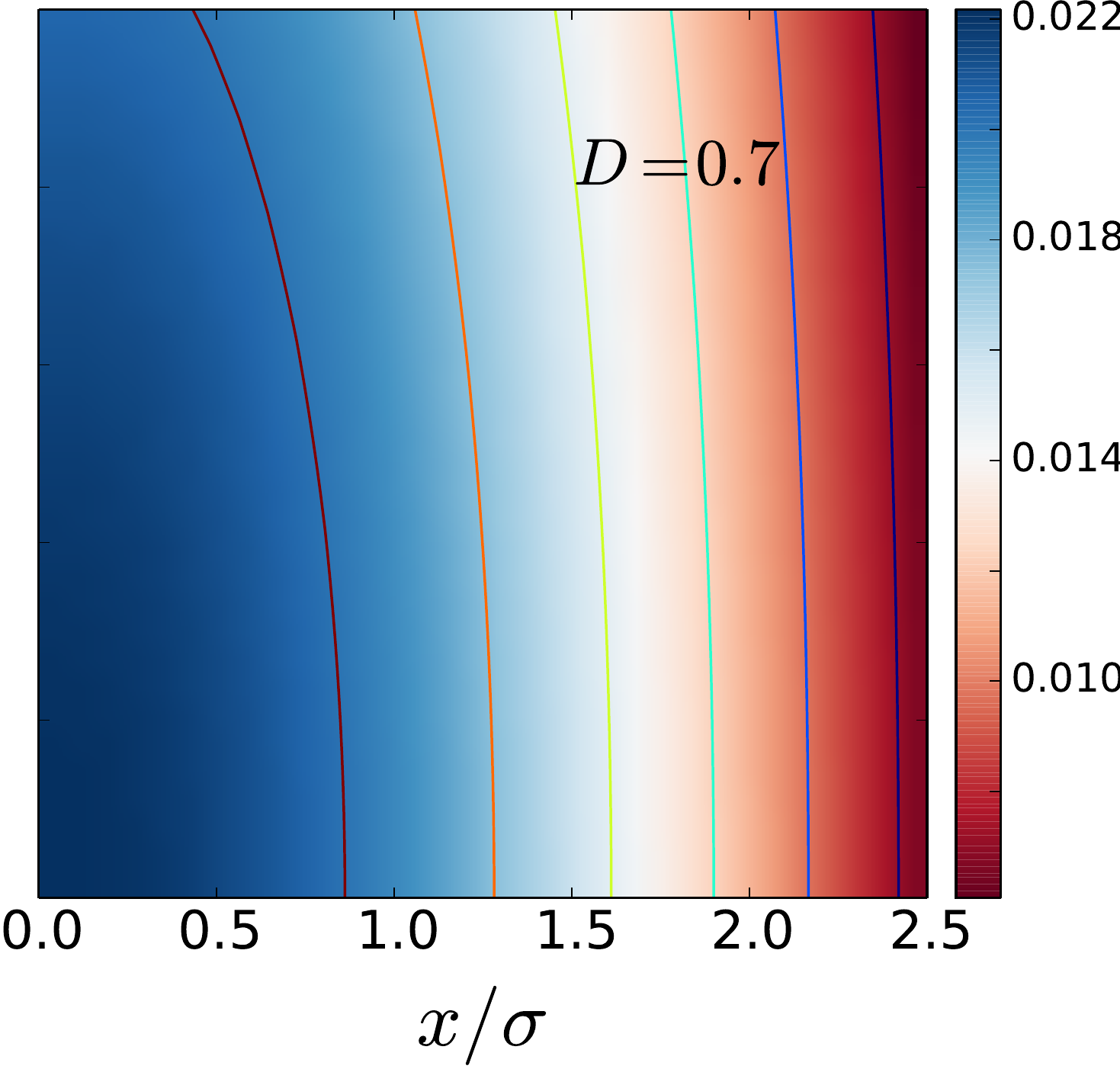} &\hspace{-0.1cm} \includegraphics[width = 0.22\textwidth]{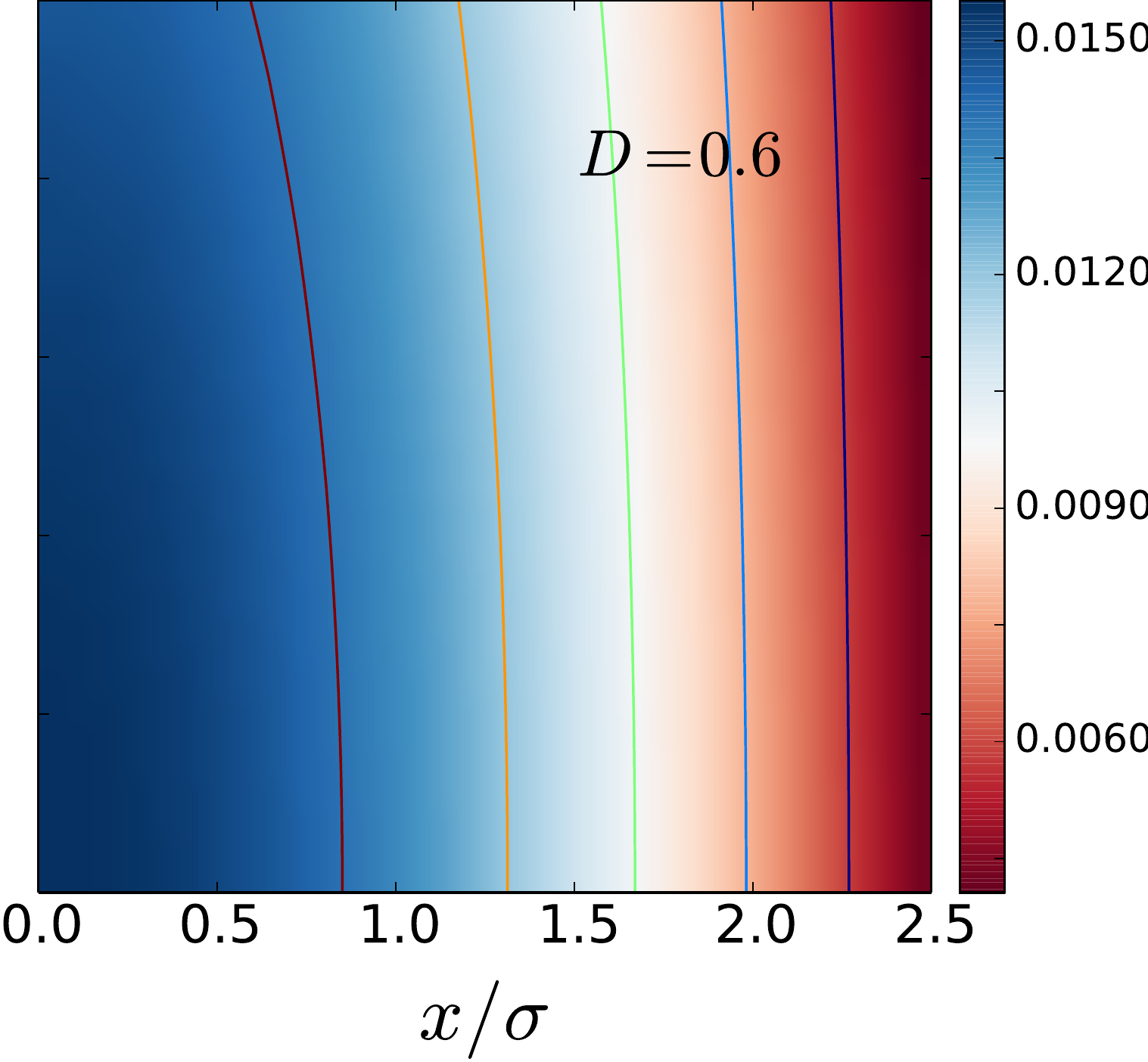} &\hspace{-0.1cm} \includegraphics[width = 0.22\textwidth]{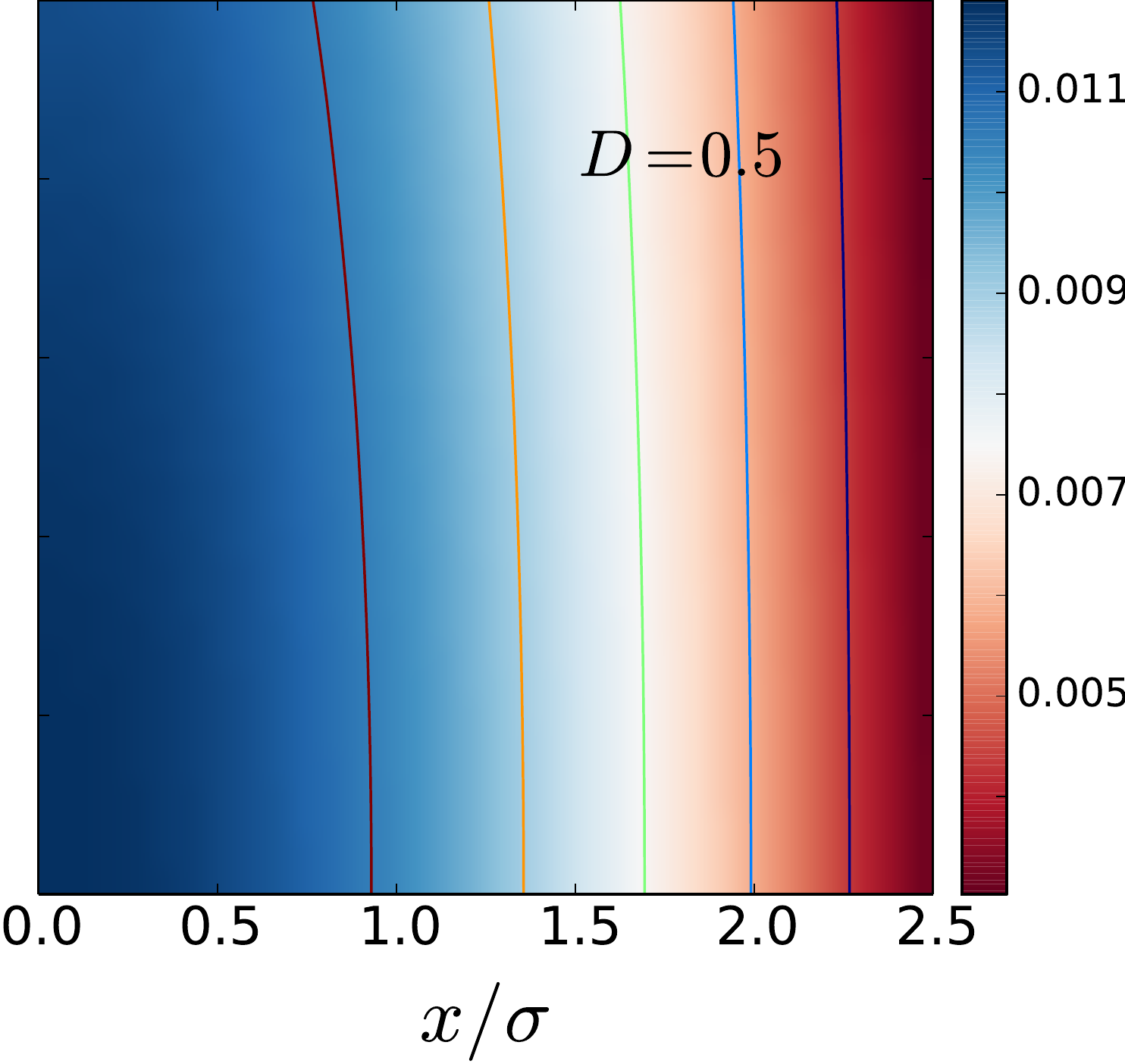}\\
  \hspace{-0.6cm} \includegraphics[width = 0.247\textwidth]{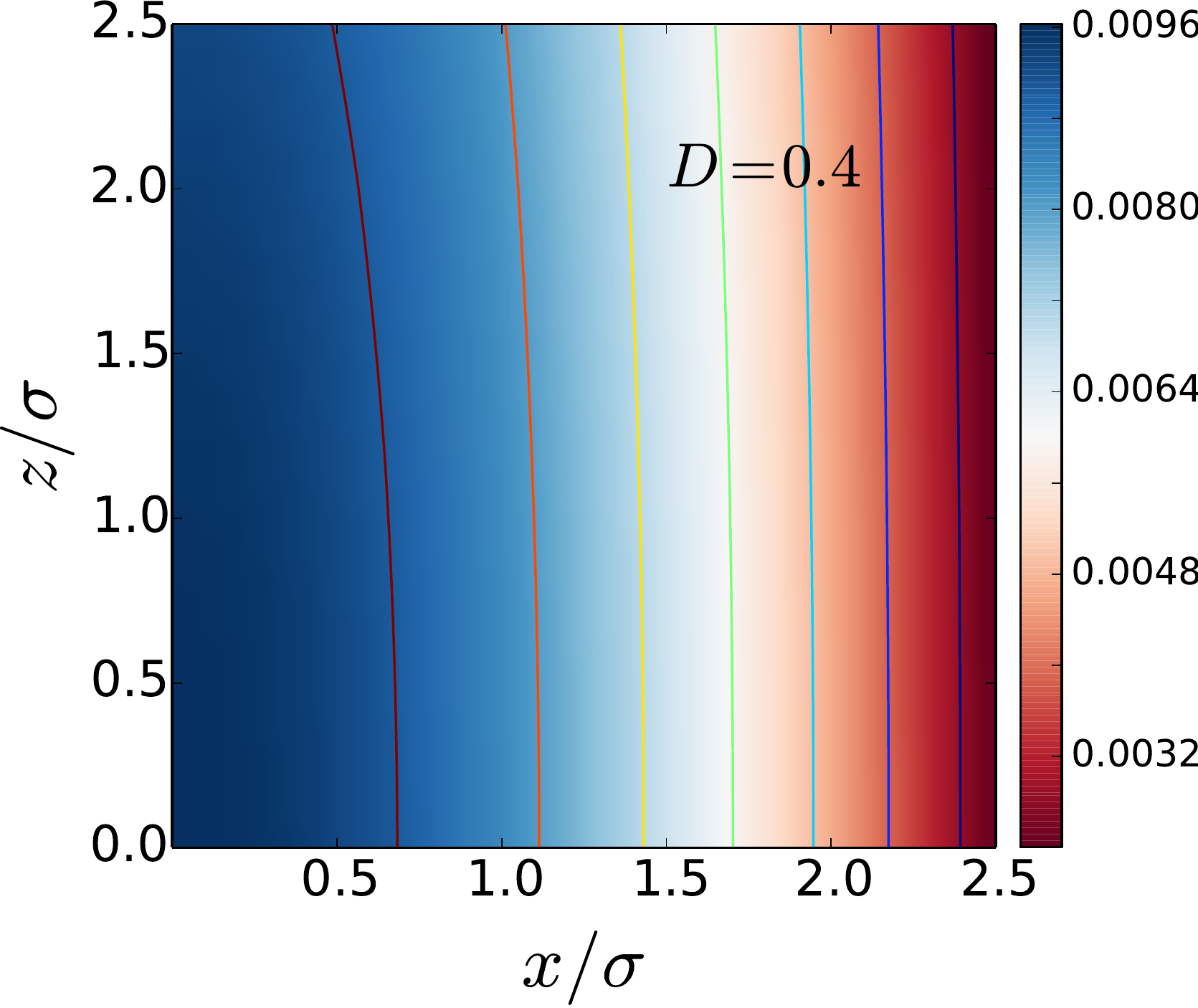} & \hspace{-0.1cm} \includegraphics[width = 0.22\textwidth]{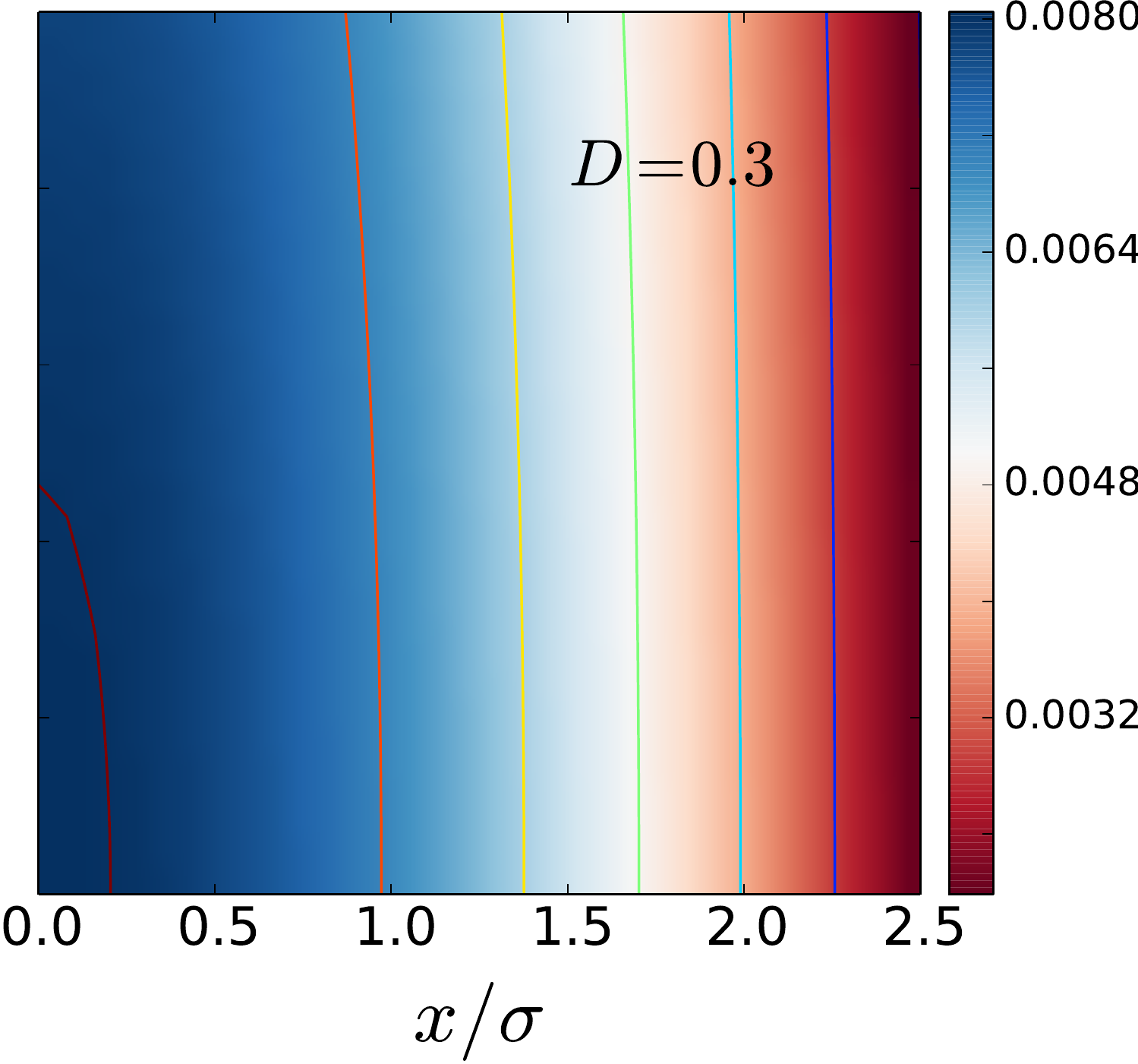} &\hspace{-0.1cm} \includegraphics[width = 0.22\textwidth]{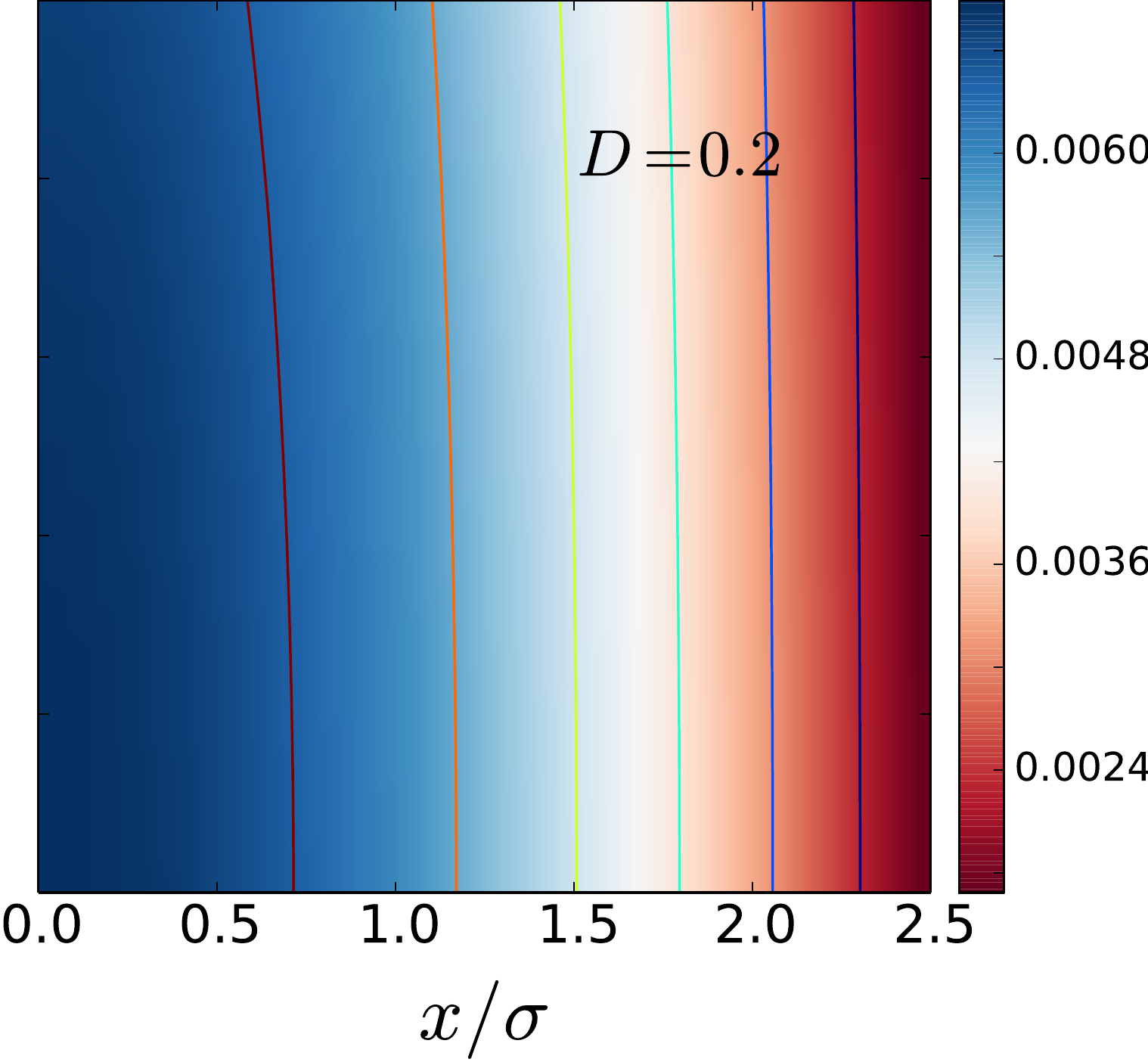} &\hspace{-0.1cm} \includegraphics[width = 0.22\textwidth]{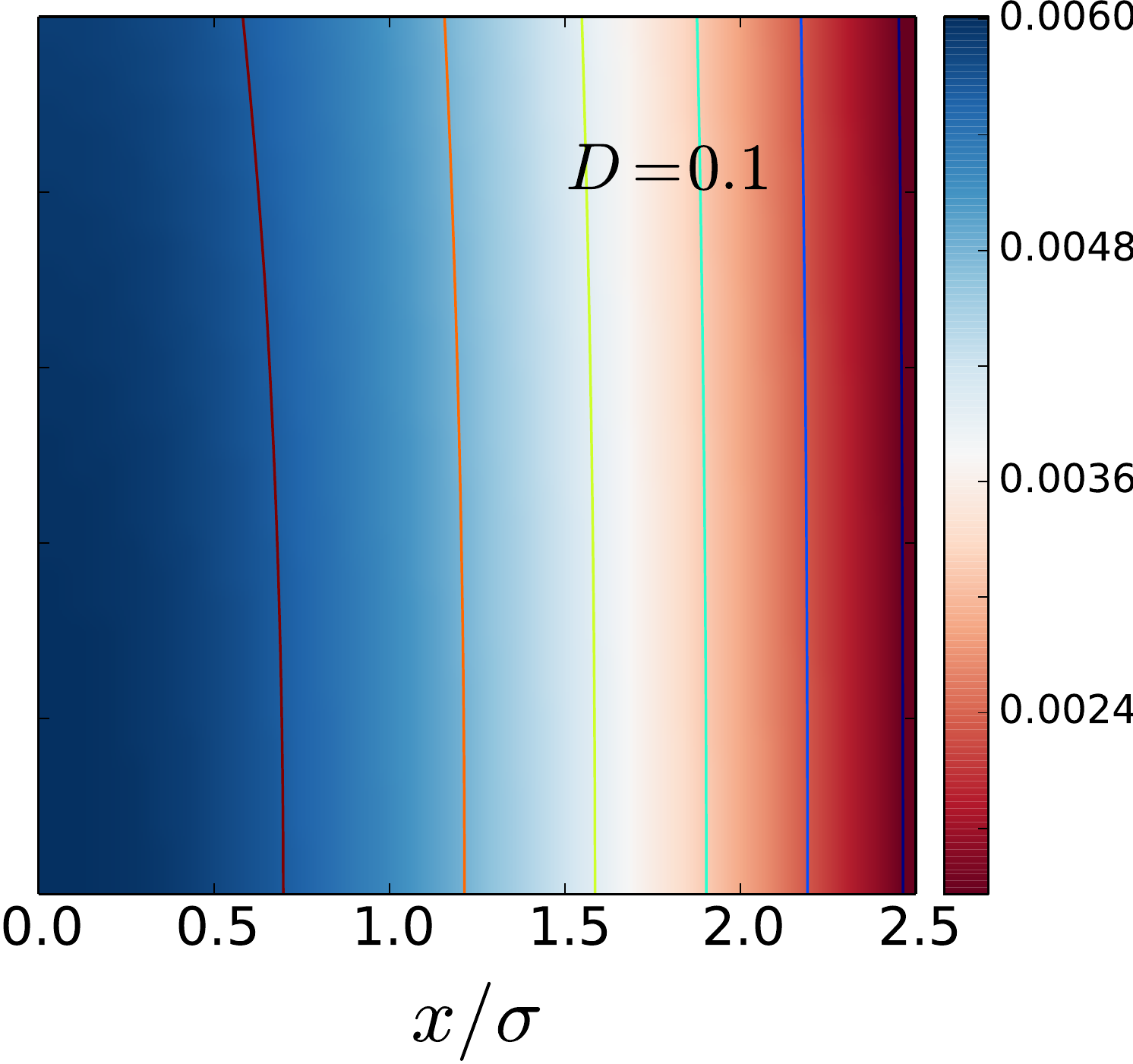}
 \end{tabular}
   \caption{Colorplots of $\tilde \rho(\rr,t) = \rho(\rr,t)-\rho_b$ shown in the $x$-$z$ plane at time $\tilde t = \sigma^2 t/D_{zz} = 69$ with $D_{zz}=1$
   for different values of $D\leq 1$ (see legend). These density distributions are the result of an inverse Fourier transformation of the results in Fig. \ref{kxkzplanes}.
   The top row of figures correspond to small anisotropy whereas the bottom row to large anisotropy.
   In the case of isotropic diffusion, $\tilde \rho(\rr,t)$ grows isotropically from the origin.
   The radially symmetrical distribution becomes practically one dimensional in the $z$-direction, even for relatively small anisotropy as can be seen in for $D=0.8$.
   Anisotropy significantly slows down the dynamics; the magnitude of the density fluctuations decreases strongly with increasing anisotropy.
 }
 \label{rhokxkzplanes}
\end{figure*}

The approach that we follow is the same as in Ref. \cite{archer2004dynamical}
We consider a homogeneous fluid that has been rapidly quenched to the region of the phase diagram inside the spinodal.  
We consider small density fluctuations $\tilde \rho(\rr,t) = \rho(\rr,t)-\rho_b$ about the bulk fluid density, $\rho_b$, and obtain an equation for the growth of these fluctuations. 
We perform a Functional Taylor expansion of the excess free energy in Eq. \eqref{functional} about the bulk density truncated at quadratic order:\cite{evans1979nature,evans1979spinodal}
\begin{align}
\F_{\mathrm{ex}}[\rho(\rr)] &= \F_{\mathrm{ex}}[\rho_b] + \int d\rr \left. \frac{\delta \F_{\mathrm{ex}}[\rho]}{\delta \rho(\rr)} \right |_{\rho_b}\tilde \rho(\rr,t)  \nonumber \\  &+ \frac{1}{2}\int d\rr \int d\rr' \left.  \frac{\delta^2\F_{\mathrm{ex}}[\rho]}{\delta \rho(\rr) \delta \rho(\rr')}\right |_{\rho_b}\tilde \rho(\rr,t)\tilde \rho(\rr',t).
\label{taylor}
\end{align}
Using Eqs. \eqref{c1} and \eqref{c2}, Eq. \eqref{taylor} can be rewritten as 
\begin{align}
\F_{\mathrm{ex}}[\rho(\rr)] &= \F_{\mathrm{ex}}[\rho_b]  + \mu_{\mathrm{ex}} \int d\rr \tilde \rho(\rr,t) \nonumber \\&-\frac{k_B T}{2}\int d\rr \int d\rr' c^{(2)}(|\rr -\rr'|)\tilde\rho(\rr,t)\tilde \rho(\rr',t),
\label{taylor2}
\end{align}
where we have used that in a bulk homogeneous fluid $\left. c^{(2)}(\rr, \rr') \right |_{\rho_b} = c^{(2)}(|\rr - \rr'|;\rho_b)$. 
Using Eq. \eqref{taylor2} to calculate the flux in Eq. \eqref{flux}, we get from the continuity equation \eqref{continuity} the following equation for the time evolution of the density
\begin{align}
\frac{\partial \rho(\rr,t)}{\partial t}  &= \nabla \cdotp \D \cdotp \rho(\rr,t) \nabla \ln[\rho(\rr,t) \Lambda^3] \\ &- \nabla \cdotp \D \cdotp \rho(\rr,t) \nabla \int d\rr'  c^{(2)}(|\rr - \rr'|) \tilde \rho(\rr',t),
\label{spinodal1}
\end{align}
where we have used that $\nabla \mu_{\mathrm{ex}}  = 0$ in a bulk homogeneous fluid. 
The equation for the density fluctuations $\tilde \rho(\rr,t)$ is obtained from Eq. \eqref{spinodal1} as
\begin{align}
\frac{\partial \tilde\rho(\rr,t)}{\partial t} &= \nabla \cdotp \D \cdotp \nabla \tilde\rho(\rr,t) \\ &- \rho_b \nabla \cdotp \D \cdotp \nabla \int d\rr'  c^{(2)}(|\rr - \rr'|) \tilde \rho(\rr',t) \nonumber \\ &- \nabla \cdotp \D \cdotp \tilde\rho(\rr,t) \nabla \int d\rr'  c^{(2)}(|\rr - \rr'|) \tilde \rho(\rr',t).
\label{spinodal2}
\end{align}

To analyze the unstable modes of the density fluctuations we consider Eq. \eqref{spinodal2} in Fourier space. 
The Fourier transformation of the density fluctuations is defined as 
$\hat \rho(\kk,t) = \int d\rr \tilde \rho(\rr,t) \exp(-i \kk \cdotp \rr)$ and of the direct pair correlation function as $\hat c^{(2)}(\kk) = \int d\rr c^{(2)}(\rr) \exp(-i \kk \cdotp \rr)$. 
With these definitions, the DDFT equation in Fourier space for the chosen quadratic approximation to $\F_{\mathrm{ex}}[\rho(\rr)]$ becomes
\begin{align}
\frac{\partial \hat \rho(\kk,t) }{\partial t} &= -(\kk \cdot \D \cdot \kk)\left(1 -\rho_b \hat c^{(2)}(\kk)\right) \hat \rho(\kk,t) \nonumber \\ &+\frac{1}{(2\pi)^3}\int d\kk' (\kk \cdot \D \cdot \kk')\hat \rho(\kk - \kk',t) \hat c^{(2)}(\kk') \hat \rho(\kk',t).
\label{fourier}
\end{align}
\noindent
The first term is linear in $\hat \rho(\kk,t)$ and governs the short-time evolution of the density fluctuations. 
The second term is nonlinear in $\hat \rho(\kk,t)$ and captures the coupling between different Fourier components of the density fluctuations. 
The corresponding equation for a system with an isotropic diffusion rate $D_0$ can be obtained as a special case of Eq. \eqref{fourier} by replacing $\kk \cdot \D \cdot \kk'$ 
with $D_0 \kk \cdot \kk'$. 
In this case, the diffusion rate $D_0$ simply appears as a scale on the right hand side of Eq. \eqref{fourier} and can be removed by redefining the time variable; this simplification is not possible in the case of an anistropic diffusion tensor. 


The short-time evolution of the density fluctuations is obtained by neglecting the second term on the right hand side of Eq. \eqref{fourier}. 
The resulting equation is a linear equation and can be solved for $\hat \rho(\kk,t)$ as
\begin{align}
\hat \rho(\kk,t)  = \hat \rho(\kk,0) \exp\left[-\frac{\kk \cdot \D \cdot \kk}{S(k)}t\right],
\label{rhoexponential}
\end{align}
where $S(k) = (1-\rho_b \hat c^{(2)}(k) )^{-1}$ and $k=|\kk|$. 
This linear theory predicts that (a) different Fourier components $\hat \rho(\kk,t)$ evolve independently and 
(b) $\hat \rho(\kk,t)$ grows exponentially in time.
For an equilibrium homogeneous fluid at a state point outside the spinodal, $S(k)$ is the well-known structure factor of the fluid.\cite{hansen1990theory} 
Since for an equilibrium fluid $S(k) >0$ for all $\kk$, it follows that for state point outside the spinodal all Fourier modes decay in time implying that the fluid is stable to small perturbations. 
However, the situation is different for state point chosen inside spinodal. 
For a chosen $\F_{\mathrm{ex}}$, $S(k)$ can become less than zero for $k< k_c$ where the value of $k_c$ depends on how deep the fluid is quenched inside the spinodal. 
This implies that the Fourier components with $k < k_c$ will grow exponentially in time. 
For a chosen $\F_{\mathrm{ex}}$, one can determine the phase diagram and chose the state point inside the spinodal. 
It is then straightforward to determine the Fourier modes that will grow exponentially.

\begin{figure}[t]
 \centering
 \vspace{-1cm}
 \includegraphics[width=\columnwidth]{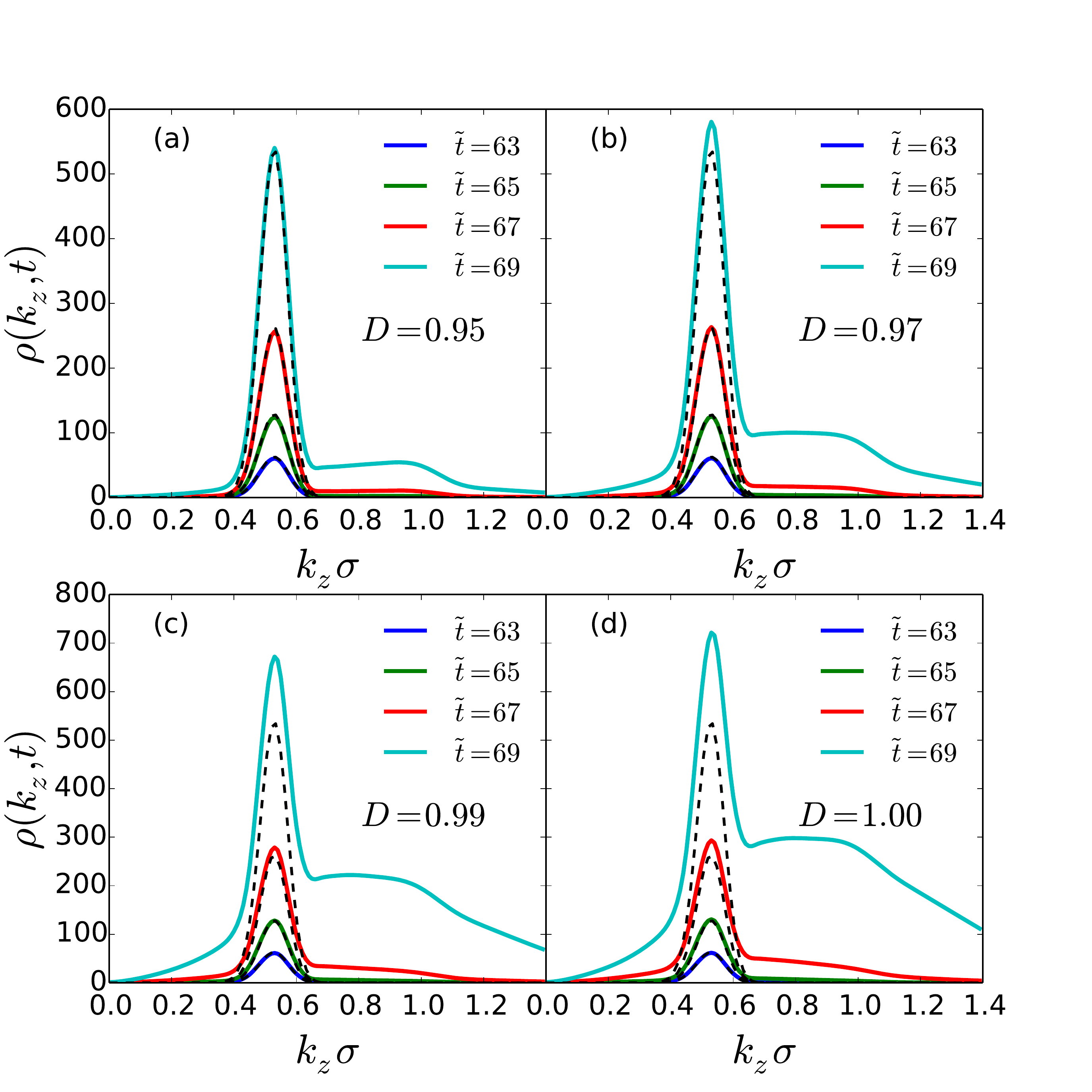}
 \caption{Fourier components of the density fluctuations $\tilde \rho(k_z,t)$ along the $k_z$-direction are shown for different times $\tilde t = \sigma^2 t/D_{zz}$ with $D_{zz}=1$. 
 The diffusion in the $x$-$y$ plane is slightly slower than in the $y$-direction with $D_{xx}=D_{yy} = 0.95$ in (a), 0.97 in (b), 0.99 in (c), 
 and the isotropic case of $D_{xx} = D_{yy} = D_{zz} = 1$ in (d). 
 Even small anisotropy significantly reduces the nonlinear contribution, i.e., the coupling of different Fourier components,
 as can be seen in the slow growth of those $k$-components 
 which lie inside the stable region ($k>k_c \approx 0.8$). 
 The dashed lines show the exponential growth predicted by the linear theory (Eq. \eqref{rhoexponential}).
 }
 \label{rhokz}
\end{figure}

\section{Results and discussion}\label{results}
It is clear from Eq. \eqref{fourier} that in order to study the dynamics of spinodal decomposition one only needs to specify the direct pair correlation function. 
At this point it becomes neccessary to assume an approximate functional for the excess Helmholtz energy which we take the same as in Ref. \cite{archer2004dynamical} 
The interested reader can find the details in Ref. \cite{archer2004dynamical} This model system exhibits fluid-fluid (gas-liquid) phase transition. 
Briefly, the system is three dimensional and is composed of particles interacting via pair potential which is infinitely 
repulsive for $|\rr|\leq \sigma$ and attractive for $|\rr| > \sigma$. The attractive part of the potential has a Yukawa form
\begin{align}
v_{at}(\rr) = -\frac{a \exp(-|\rr|/\sigma)}{4\pi \sigma^2 |\rr|},
\label{vat}
\end{align}
where $a$ is a positive parameter that determines the strength of the attraction. 
Treating the attractive interactions in a mean-field fashion, the following expression is obtained for the direct pair correlation function in Fourier space:
\begin{align}
\hat c^{(2)}(k) = \hat c_{\mathrm{PY}}(k) + \frac{\beta a}{1 + (k\sigma)^2},
\end{align}
where $\hat c_{\mathrm{PY}}(k)$ is the Percus-Yevick approximation for the hard-sphere pair direct correlation function. \cite{hansen1990theory}

The phase diagram of the fluid system under consideration is shown in Fig. \ref{phasediagram}(a). All the results below are shown for the quench corresponding to sudden cooling from state point $A$ outside the spinodal to the point $B$ inside the spinodal. In Fig. \ref{phasediagram}(b) we plot the function $-k^2/S(k) \equiv -k^2(1-\rho_b \hat c^{(2)}(k))$ which is the argument of the exponential in Eq. \eqref{rhoexponential} corresponding to the special case of isotropic diffusion with $\D=\mathbb{1}$. As can be seen in Fig. \ref{phasediagram}(b), this function is positive for the wavenumbers $k<k_c$ where $k_c$ is determined by the condition $\rho_b\hat c^{(2)}(k_c) =1$.

\begin{figure}[t]
 \centering
  \vspace{-1cm}
 \includegraphics[width=\columnwidth]{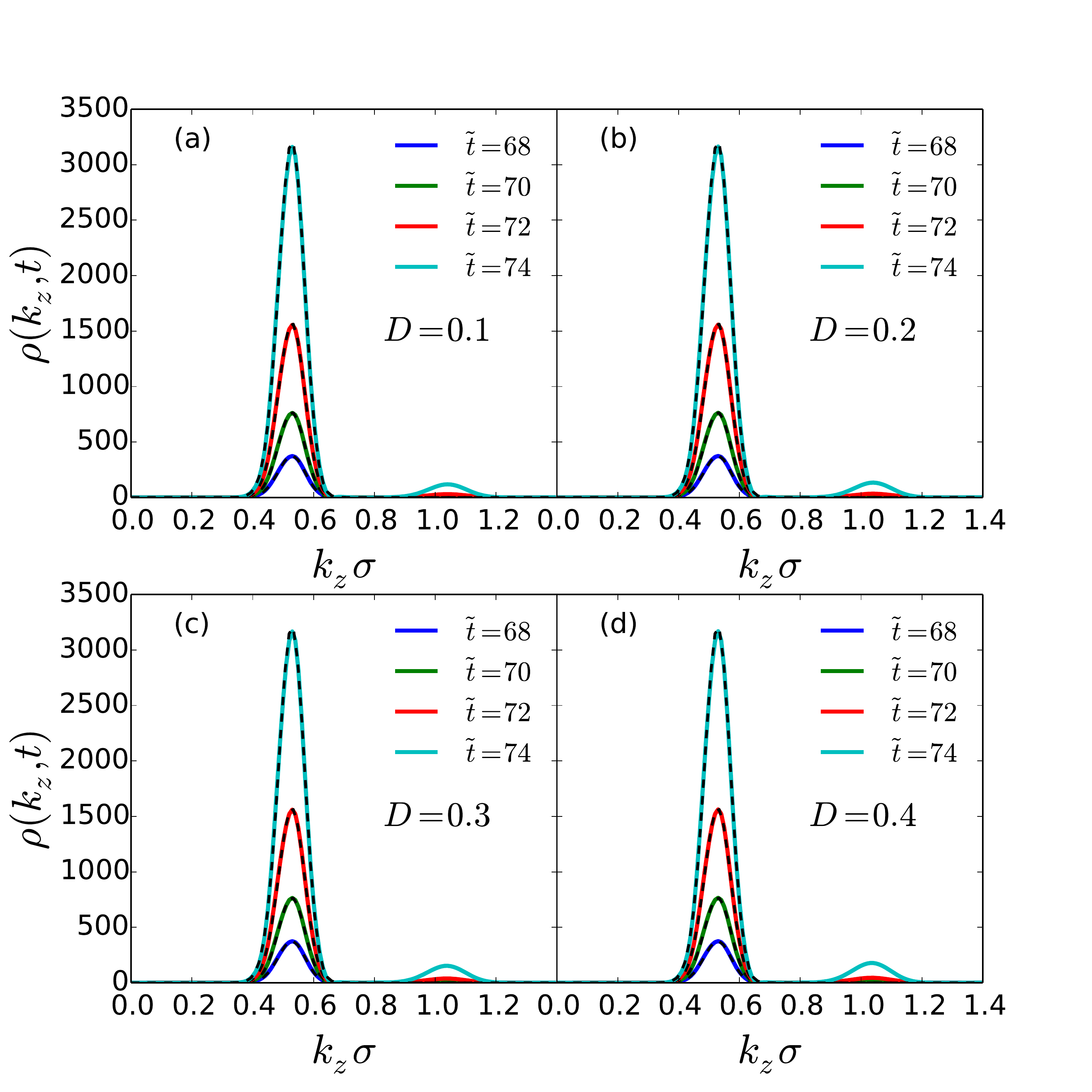}
 \caption{Same as in Fig. \ref{rhokz}, but for longer times and larger anisotropy, $D_{xx}=D_{yy} = 0.1$ in (a), 0.2 in (b), 0.3 in (c), and 0.4 in (d). 
 Anisotropic diffusion strongly reduces the growth of $k_z$-modes  in the $k_z>k_c$ region. 
 This is apparent in the relatively smaller second peak located at $k_z \sigma \approx 1.1$. 
 The dashed lines show the exponential growth predicted by the linear theory (Eq. \eqref{rhoexponential}). 
 The agreement with the predictions of the linear theory is nearly perfect implying that the nonlinear contribution to the growth of density fluctuations is negligible.
 }
 \label{rhokz_strong_anisotropy}
\end{figure}

\begin{figure*}[t]
 \centering
  \vspace{-1cm}
 \begin{tabular}{cc}

  \includegraphics[width=\columnwidth]{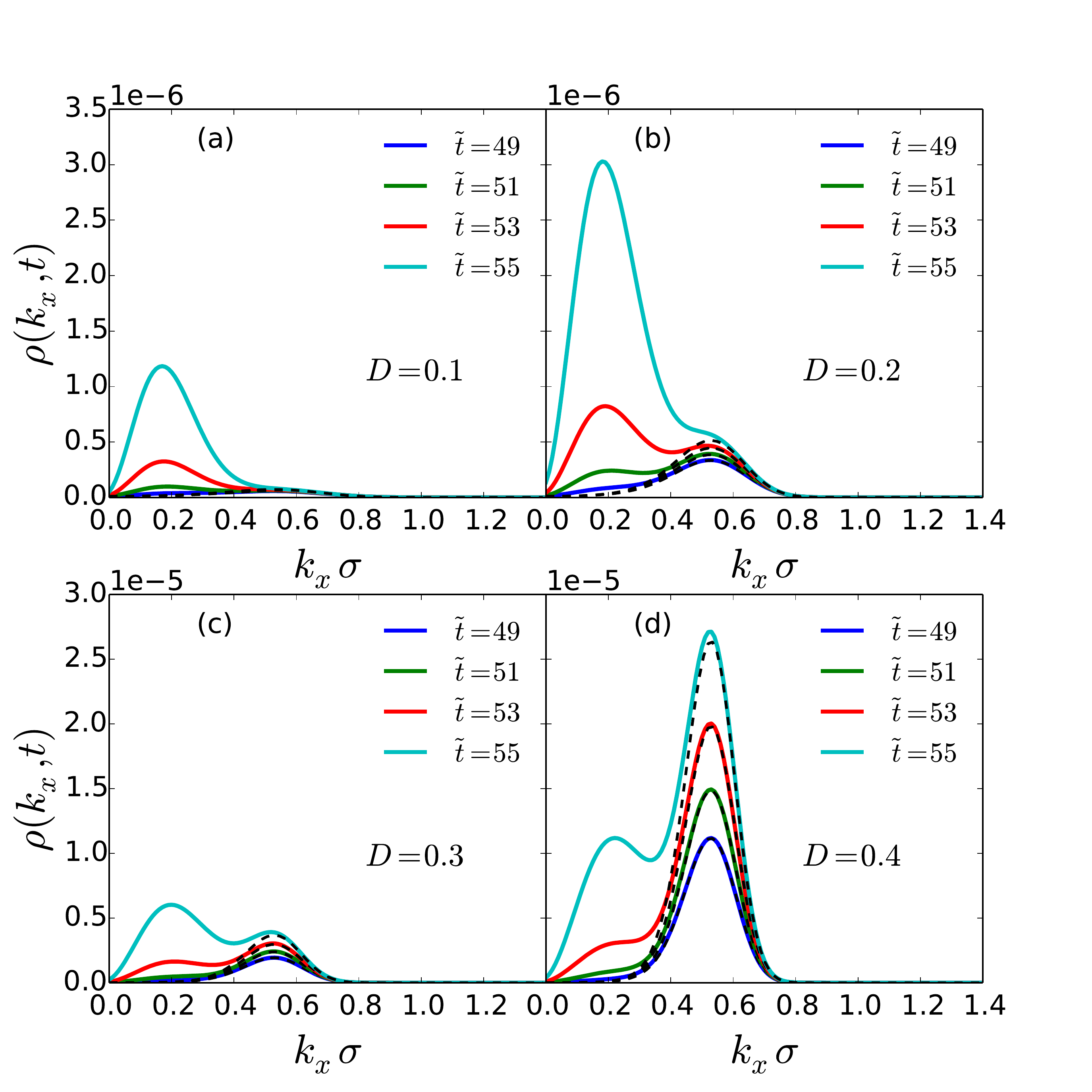} &
   \hspace{-1cm} \includegraphics[width=\columnwidth]{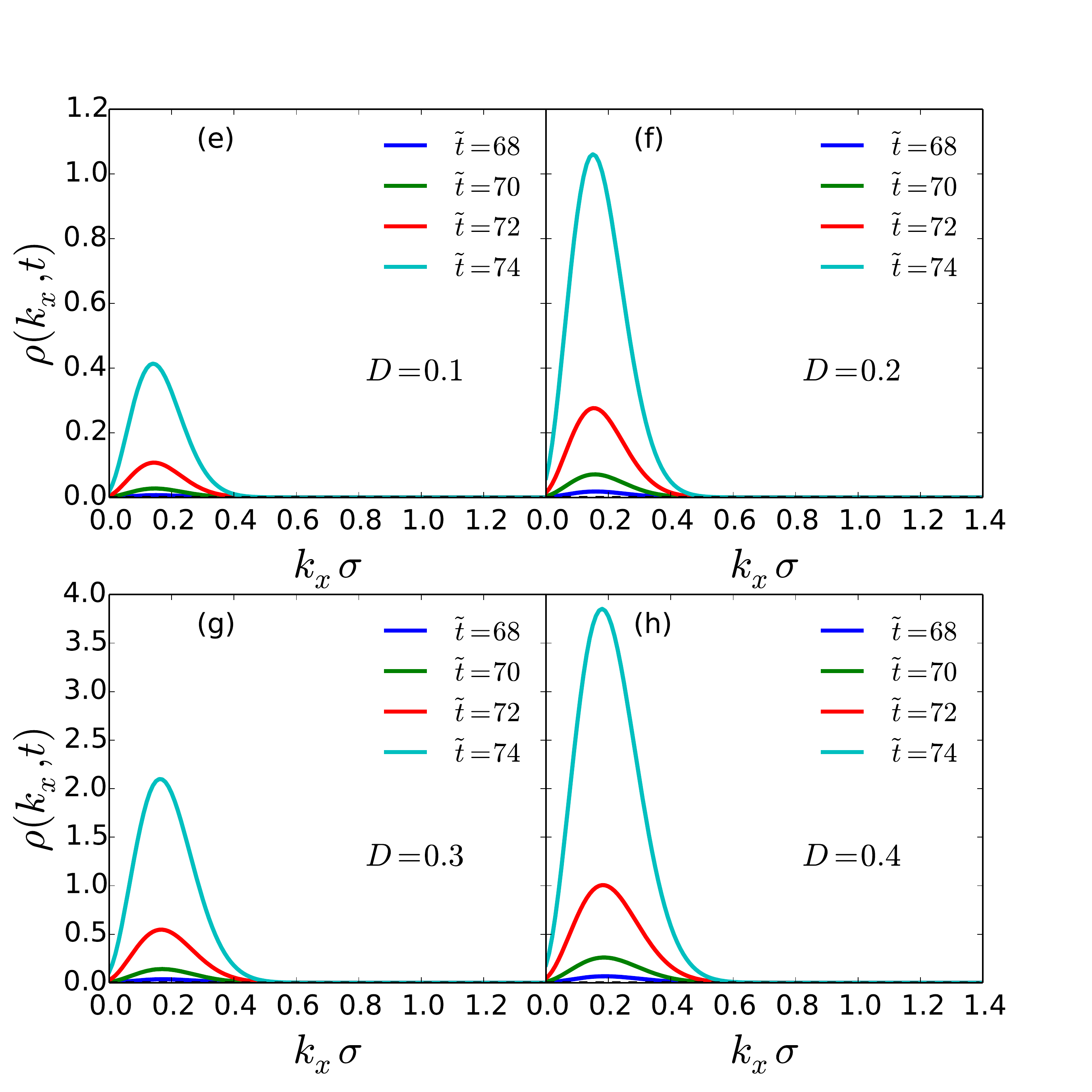}
 \end{tabular}
   \caption{Fourier components of the density fluctuations $\tilde \rho(k_x,t)$ along the $(k_x,0,0)$ direction is shown for times $\tilde t = \sigma^2 t/D_{zz}$ with $D_{zz}=1$ The diffusion in the $x$-$y$ plane is much slower with $D_{xx}=D_{yy} = 0.1$ in (a,e), 0.2 in (b,f), 0.3 in (c,g), and 0.4 in (d,h). 
   Figures in the left column (a-d) show the Fourier components for times $\tilde t = 49$, 51, 53, and 55 whereas the 
   figures in the right column (e-h) show the same for times $\tilde t =68$, 70, 72, and 74. Note that the scale of y-axis in (a-d) is much smaller than in (e-h). 
   Acording to the linear instability analysis the fastest growing Fourier mode is the one with $k_x\sigma \approx 0.5$ which should grow exponentially at short times. 
   The dashed lines show the predicted exponential growth from the linear theory (Eq. \eqref{rhoexponential}). 
   In (a), which corresponds to the smallest value of $D_{xx}$, there is no peak at $k_x\sigma \approx 0.5$. 
   The peak becomes increasingly visible in going from (a) to (d).
   The growth of density fluctuations in the $x$-$y$ plane is predominantly driven by the nonlinear contributions coming from the coupling of different Fourier components. 
   This is evident in the figures (e-h), where after sufficiently long time, there is only a single peak located at $k_x \sigma \approx 0.17$.
 }
 \label{rhokxy}
\end{figure*}

\begin{figure}[t]
 \centering
 \vspace{-1cm}
 \includegraphics[width=\columnwidth]{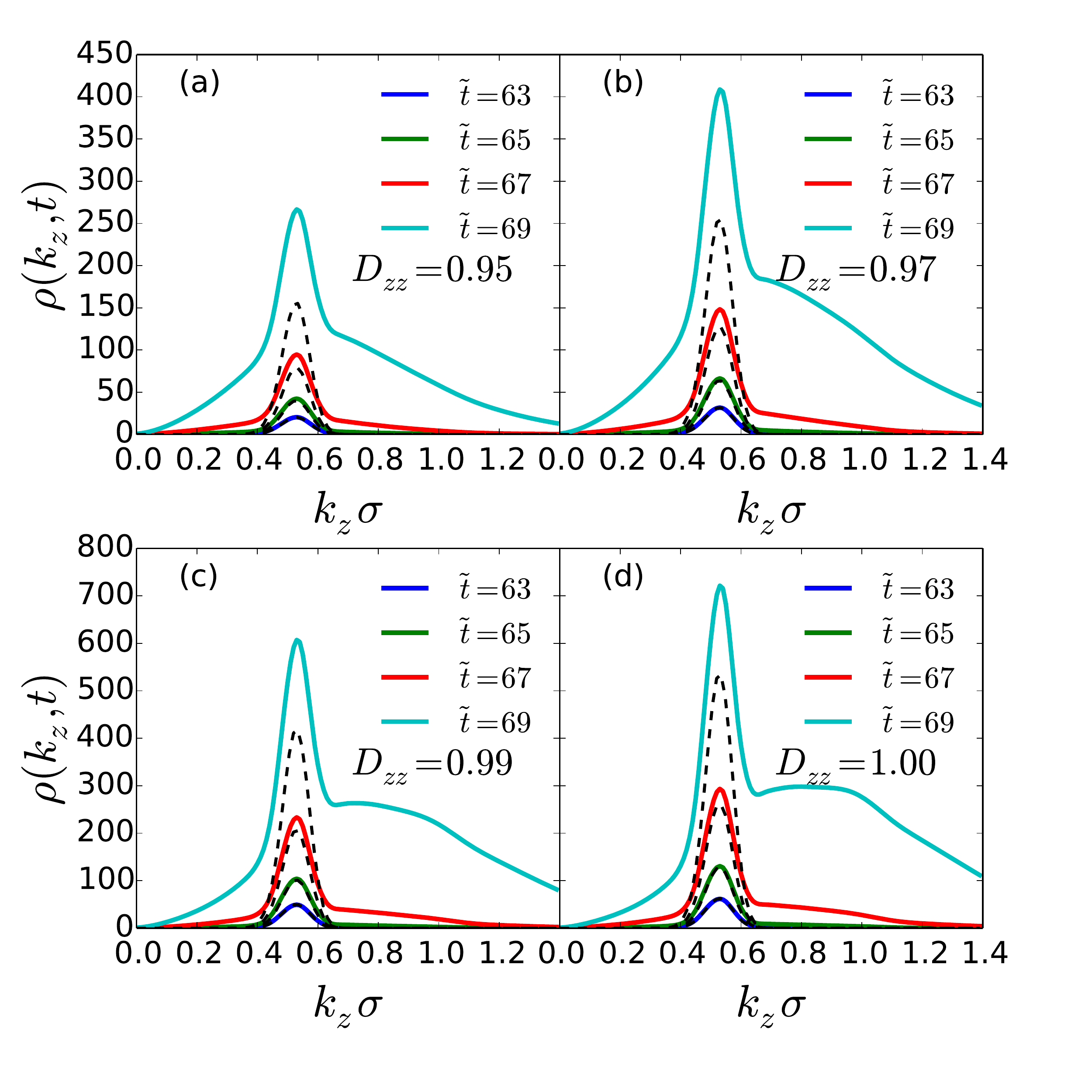}
 \caption{Fourier components of the density fluctuations $\rho(k_z,t)$ along the $k_z$-direction is shown for times $\tilde t = \sigma^2 t/D_{xx}$ with $D_{xx} = 1$. The diffusion in the $z$-direction is slightly slower with $D_{zz}= 0.95$ in (a), 0.97 in (b), 0.99 in (c), and 1 in (d). 
 Though anisotropic diffusion reduces the growth of $k_z$-modes outside the $k_z < k_c$ region, as in Fig. \ref{rhokz}, the effect is comparatively weaker.
 The dashed lines show the predicted exponential growth from the linear theory (Eq. \eqref{rhoexponential}).
 }
 \label{rhokz_zslow}
\end{figure}

%
%

To model anisotropic diffusion, we consider the following diagonal diffusion tensor:
\begin{align}
\D = \left(\begin{array}{ccc} D_{xx}=D & 0 & 0\\ 0 & D_{yy}=D & 0\\ 0 & 0 & D_{zz}  \end{array}\right),
\end{align}
where $D$ and $D_{zz}$ are parameters. Note that such tensor is cylindrically symmetric.

We use the approximate form for $\hat c^{(2)}(k)$ as input to Eq. \eqref{fourier} and calculate the time evolution of density fluctuations for the state point $k_B T \sigma^3/a = 0.05$ 
and $\eta= 0.2$ (See Fig. \ref{phasediagram}). 
This state point lies well inside the spinodal region. We assume that at time $t=0$  all the Fourier components have the value $\hat \rho(\kk,0) = 10^{-8}$. 
In real space this corresponds to a single perturbation at the origin in the otherwise uniform density profile.
The cylindirical symmetry of the diffusion tensor together with the chosen intial conditions imply that $\tilde \rho(\kk,t)$ is completely described by $\tilde \rho(\{k_x,0,k_z\},t)$. 
In Fig. \ref{kxkzplanes}, we show $\tilde \rho(\{k_x,0,k_z\})$ at time $\tilde t = \sigma^2 t/D_{zz} = 69$ for different values of $D$. 
Diffusion along the $z$-direction is fixed to $D_{zz} = 1$. 
The choice of $\tilde t = 69$ corresponds to intermediate-time dynamics of spinodal decomposition in an isotropic system ($\D=\mathbb{1}$). 
In the case of isotropic diffusion $\tilde \rho(\{k_x,0,k_z\})$ has circular contours in the $k_x$-$k_z$ plane. 
The Fourier components with wavenumbers $k<k_c$ exhibit strong growth with the fastest growing mode at $k\sigma \approx 0.5$. 
Fourier components with wavenumbers outside the instability region also grow due to the nonlinear coupling of different Fourier components. 
It is clear from the figure \ref{kxkzplanes} that anisotropy strongly impacts the nonlinear coupling. The growth along the 'slow' directions is strongly suppressed. 
The suppression seems to be significant even for relatively small anisotropy ($0.8<D<1$). For large anisotropy ($D<0.5$), only the Fourier components along the $(0,0,k_z)$ direction show 
any significant growth. In Fig. \ref{rhokxkzplanes}, we plot the real-space density (fluctuation) distribution obtained from inverting the Fourier transform of Fig. \ref{kxkzplanes}. 
The density distribution is spherical symmetrical for $D=\mathbb{1}$. 
The radially symmetrical distribution becomes practically one-dimensional in the $z$-direction even for relatively small anisotropy as can be seen in Fig. \ref{kxkzplanes} for $D=0.8$. 
What is more interesting is how anisotropy slows down the dynamics; at a given instant of time, the magnitude of the density fluctuations is much smaller than that of the isotropic diffusion case.

In order to better understand the dynamics at different times, we focus on the time evolution of Fourier components $\tilde \rho(k_z,t)$ along the $(0,0,k_z)$ direction, which we refer to as the $k_z$-direction. 
Although a complete description of the decompostion dynamics can only be obtained by considering the full $\tilde \rho(\kk,t)$, 
by selectively focusing on the dynamics along $k_z$-direction, one can clearly see the effect of anisotropy on the intermediate-time dynamics of spinodal decomposition. 

We first consider the case of small anisotropy. 
The diffusion along the $z$-axis is fastest with $D_{zz} = 1$ and the diffusion in the $x$-$y$ plane is slightly slower with $D_{xx} = D_{yy} = D$. 
In Fig. \ref{rhokz} we plot the Fourier components of the density fluctuation along the $k_z$-direction for different values of $D$. 
We also show the results ignoring the nonlinear contribution. On neglecting the nonlinear contribution, only the $k_z$-components which lie within the range $k_z <k_c\approx 0.8$, 
for which $S(k_z) <0$, grow exponentially. The growth of $k_z$-components outside this range is due to the coupling of Fourier components. 
For short times, only the components inside the linear instability region grow. As these modes lying within $k_z <k_c$ grow, they couple to feed the growth of $k_z$ modes lying outisde 
the instability region. It is for the growth of these $k_z$ modes lying outside the $k_z <k_c$ region that the effect of anisotropy is most pronounced. 
As can be seen in Fig. \ref{rhokz}, the nonlinear contribution to the density fluctuations is significantly reduced in the case of anisotropic diffusion. 
Surprisingly, even a relatively small anisotropy can significantly reduce the coupling of different Fourier components.

In case of strongly anisotropic diffusion ($D<0.5$), the nonlinear contribution to the growth of the density fluctuations is highly supressed. 
We show in Fig. \ref{rhokz_strong_anisotropy} the Fourier components of the density fluctuations along the $k_z$-direction for small values of $D$. 
The diffusion along the $z$-axis remains the fastest. As can be seen in Fig. \ref{rhokz_strong_anisotropy}, there is an additional small peak outside the $k_z <k_c$  
region which grows slowly in time. The height of the secondary peak remains negligible in comparison to the main peak for the times considered. The growth of the main peak is accurately described by the linear theory. 
It follows that the nonlinear contribution to the growth of density fluctuations within the $k_z <k_c$ is negligible from the Fourier components with $k_z>k_c$. 
Of course, if one considers the density fluctuations on longer time scales than those shown in Fig. \ref{rhokz_strong_anisotropy}, the second peak will grow larger. 
However, in this case the density fluctuations (for instance in the main peak) will become so large that the Taylor expansion of the Helmholtz free energy functional in Eq. \eqref{taylor} 
up to quadratic order is insufficient to describe the spinodal decomposition at intermediate times. 

On considering the dynamics in the $k_x$-$k_y$ plane, one finds that the growth of the density fluctuations is predominantly driven by the nonlinear contributions. In Fig. \ref{rhokxy} we plot the Fourier components of the density fluctuations along the $k_x$ direction for same values of $D_{xx}=D_{yy}$ as in Fig. \ref{rhokz_strong_anisotropy}. It is useful to first consider the extreme case of $D_{xx}=0$. In this case, the density fluctuations of the Fourier component with a given wave number $k_x$ can grow only due to the coupling of other Fourier components. Not surprisingly, this implies a complete absence of the linear regime. 
When $D_{xx}$ is slightly greater than zero, there will be linear regime which will persist on a time scale governed by $D_{xx}$ and $k_x$, as can be seen in Fig. \ref{rhokxy}. 
For $D_{xx} = 0.1$ there is no peak at the location $k_x \sigma \approx 0.5$ for the times considered in Fig. \ref{rhokxy}(a), as predicted by the linear theory.
On increasing $D_{xx}$ the peak becomes increasingly visible at the location $k_x \sigma \approx 0.5$ (see Figs. \ref{rhokxy}(b-d)). For all $D_{xx}$ there is an additional peak that grows at the location $k_x \sigma\approx 0.17$. With increasing time, the nonlinear contribution becomes dominant and there remains a single peak at $k_x \sigma\approx 0.17$ (see Figs. \ref{rhokxy}(e-h)).

Until now we have considered anisotropy of the form such that the diffusion is fastest along the $z$-direction. With increasing anisotropy, the system can be considered to becomes effectively one dimensional. 
We now consider another scenario in which we fix $D_{xx} = D_{yy} = 1$ and decrease the diffusion coefficient along the $z$ direction below 1. 
The extreme case in this scenario of $D_{zz} =0$ would then correspond to an effectively two-dimensional system. We expect that in this scenario, the anisotropic diffusion would have a weaker effect than the previous scenario. 
In Fig. \ref{rhokz_zslow} we plot the Fourier components of the density fluctuation along the $z$-direction for different values of $D_{zz}$. As can be seen in this figure the anisotropy has a much weaker effect than what was observed in Fig. \ref{rhokz}. 
This can be qualitatively understood considering that in Fig. \ref{rhokz}, the anisotropy reduces the contribution to the coupling from $k_x$-$k_y$ plane whereas in Fig. \ref{rhokz_zslow} only the contribution from $k_z$-direction is reduced.

\section{Conclusions}\label{conclusions}
We studied  spinodal decomposition in a colloidal system composed of particles which diffuse anisotropically in space. 
We used DDFT to model the dynamics of spinodal decomposition at short and intermediate times. We found that spatial anisotropy significantly alters the dynamics of spinodal decomposition. 
At short times, the main effect of anisotropic diffusion is a trivial modification of the growth rate of density fluctuations. 
For later stages of decomposition, anisotropy can significantly reduce the coupling of different Fourier components. 
As a consequence of which the growth of density fluctuations at wave numbers that lie inside the range of linear stability is significantly slowed down. 
We found that the nonlinear contribution to the density fluctuations is highly sensitive to the degree of anisotropy. 
The slow growth of Fourier components outside the range $k < k_c$ is significant even for relatively small anisotropy. 
In the case of strongly anisotropic diffusion, the growth of Fourier components outside the range  $k < k_c$  is strongly suppressed. 

The results presented in this paper show that the dynamics of spinodal decomposition become quite rich when the diffusion is assumed to be anisotropic. 
However, there are some aspects that we have not addressed in this paper. For instance, consider the case of strongly anisotropic diffusion (Fig. \ref{rhokz_strong_anisotropy}). 
The secondary peak that is located at $k_z \sigma \approx 1.1$ is well seperated from the main peak located at $k_z\sigma \approx 0.5$. It is natural to ask why the $k_z$ components between the two locations do not grow for the times considered in Fig. \ref{rhokz_strong_anisotropy}. 
Another aspect that we have not addressed is the quantification of the intermediate time. 
This is particularly relevant in the case of strongly anisotropic diffusion because of the possibility of formation of sharp interfaces. 
These aspects require more mathematical analysis than presented in this paper and will be investigated in the future.

In this paper we focused on a fluid system that exhibits fluid-fluid phase transition. 
There are other dynamical phenomena which are sensitive to spatial anisotropy of diffusion. 
For instance, a colloidal system that exhibits freezing transition. In such a system anisotropic diffusion is expected to have important consequences for dynamical pattern formation during 
the freezing transition. \cite{archer2014solidification} 
Another interesting phenomenon is the laning transition in a sheared colloidal system.\cite{ackerson1988shear,chakrabarti2003dynamical,brader2011density}
It will be very interesting to study these phenomena using DDFT together with anisotropic diffusion. Anisotropic diffusion can be obtained in practise in a system of interacting colloidal 
particles subjected to Lorentz forces. It will be very interesting to perform Brownian dynamics simulations of such systems and study the above mentioned dynamical phenomena.

\section*{Conflicts of interest}
There are no conflicts to declare.





\bibliographystyle{apsrev}
%
%

\end{document}